%% file: microstructure.tex
\documentclass[11pt]{My_preprint}

\title{
    Buoyancy driven motion of non-coalescing inertial drops: microstructure modeling with nearest particle statistics. 
}

\author[1,2]{Nicolas Fintzi}
\author[1]{Jean-Lou Pierson}
\author[2]{St\'ephane Popinet}
\affil[1]{IFP Energies Nouvelles, Rond-point de l’echangeur de Solaize, 69360 Solaize}
\affil[2]{Sorbonne Universit\'e, Institut Jean le Rond d'Alembert, 4 place Jussieu, 75252 PARIS CEDEX 05, France}
\normalmarginpar

\begin{document}

\maketitle

\begin{abstract}
    In this study, we analyze the various arrangements that droplets can form within dispersed buoyant emulsions, which we refer to as the study of microstructure. 
    To this end, we have developed a novel algorithm that effectively prevents numerical coalescence between drops while maintaining a reasonable computational cost. 
    This algorithm is integrated into the Volume of Fluid (VoF) method and implemented using the open-source code \href{http://basilisk.fr}{http://basilisk.fr}. 
    Subsequently, we perform Direct Numerical Simulations (DNS) of statistically steady state mono-disperse buoyant emulsion over a broad range of dimensionless parameters, including the particle volume fraction ($\phi$), the Galileo number ($Ga$) and the viscosity ratio ($\lambda$). 
    We make use of nearest particle statistics to quantify the microstructure properties.  
    As predicted by \citet{zhang2023evolution}, it is demonstrated that the second moment of the nearest particle pair distribution can effectively quantify microstructural features such as particle clusters and layers. 
    Specifically, the findings are: (1) In moderately inertial flows ($Ga=10$), droplets form isotropic clusters. 
    In high inertial regimes ($Ga=100$), non-isotropic clusters, such as horizontal layers, are more likely to form. 
    (3) The viscosity ratio plays a significant role in determining the microstructure, with droplets that are less viscous or equally viscous as the surrounding fluid tending to form layers preferentially. 
    Overall, our study provides a quantitative measure of the microstructure 
     in terms of $Ga$, $\phi$ and $\lambda$. 
\end{abstract}

\section{Introduction}
\input{microstructure/Introduction.tex}

\section{Numerical methodology}
\label{sec:methodo}

\input{microstructure/Problem_statement.tex}
\input{microstructure/Simulation_Setup.tex}

\input{microstructure/no_coalesence.tex}

\section{Nearest particle statistics}
\label{sec:nearest}
\input{microstructure/The_nearest_particle_statistics.tex}

\section{Microstructure}
\label{sec:microstructure}
\input{microstructure/Non_Inertial.tex}

\input{microstructure/Inertial.tex}

\section{Conclusion}
\label{sec:conclusion}
\input{microstructure/conclusion.tex}

\appendix

\input{microstructure/AgeAP.tex}

\section{Numerical validations}
\label{ap:validation}
\input{microstructure/validation.tex}

\bibliography{Bib/bib_bulles.bib}

\end{document}

%% file: microstructure/Introduction.tex
Buoyancy-driven droplet flows are commonly encountered in chemical engineering processes, including gravity separators and liquid-liquid extractors. Typically, these systems are modeled using the averaged Navier-Stokes equations as described by \citep{castellano2019} for instance. However, these methods necessitate closure laws and a thorough understanding of particle pair statistics \citep{simonin1996}. The accuracy of these closure laws significantly depends on the physical properties of the fluids and the arrangement of particles, referred to as the microstructure. 
For example, in the Stokes flow regime for a dilute ordered array consisting of a periodic arrangement of spherical inclusions, the dimensionless relative velocity of the suspension decreases proportionally to the cube root of the volume fraction. Conversely, in the case of a random array of freely moving particles, the dimensionless velocities decrease linearly with the volume fraction \citep{saffman1973}.
In inertial regimes, \citet{yin2007} demonstrate that the settling velocity of randomly arranged solid spherical particles can be characterized by a power-law function of $(1-\phi)$, where $\phi$ denotes the particle volume fraction.
However, this power-law relationship fails to accurately describe the observed data for anisotropic bubble microstructures, as shown by \citet{yin2008lattice} and \citet{loisy2017}.
Additionally, \citet{cartellier2009induced} conducted experiments of buoyant bubbly flows. 
They measured the mean velocity agitation of the continuous phase for various $\phi$, 
and they demonstrated that the change in microstructure influence the scaling of agitation with the volume fractions. 
Thus, the mean drag force or mean drift velocity as well as the continuous phase agitation of suspensions, which are essential parameters in the modeling of two-phase flow problems, is influenced by the geometry of the microstructure.

Another relevant illustration involves the presence of poly-disperse distribution in chemical engineering processes. In these scenarios, Population Balance Equations (PBE) are employed to depict the distribution of droplet sizes \citep{randolph2012theory}. PBE relies on a coalescence kernel, which acts as a source term delineating the rate at which droplets merge. This kernel depends on the microstructure and the relative motion of particle pairs \citep{chesters1991modelling}. Consequently, having statistical information on particle pairs becomes essential in PBE models. This study aims to analyze the microstructure of an emulsion and the relative motion of particle pairs across a wide range of dimensionless parameters.

\begin{figure}
    \hfill
\begin{tikzpicture}
    \draw[->] (-1,2.5)--++(0,-2)node[midway, left]{$\textbf{g}$};
    \foreach \i in {1,...,5} {
    \pgfmathsetmacro{\x}{rnd}
    \pgfmathsetmacro{\y}{rnd}
    \draw[fill=gray!50] ($(\x,\y)$) circle (0.1);
    }
    \foreach \i in {1,...,5} {
    \pgfmathsetmacro{\x}{rnd}
    \pgfmathsetmacro{\y}{rnd}
    \draw[fill=gray!50] ($(\x+1,\y)$) circle (0.1);
    }
    \foreach \i in {1,...,5} {
    \pgfmathsetmacro{\x}{rnd}
    \pgfmathsetmacro{\y}{rnd}
    \draw[fill=gray!50] ($(\x+2,\y)$) circle (0.1);
    }
    \foreach \i in {1,...,5} {
        \pgfmathsetmacro{\x}{rnd}
        \pgfmathsetmacro{\y}{rnd}
        \draw[fill=gray!50] ($(\x+1,\y+1)$) circle (0.1);
    }
    \foreach \i in {1,...,5} {
    \pgfmathsetmacro{\x}{rnd}
    \pgfmathsetmacro{\y}{rnd}
    \draw[fill=gray!50] ($(\x+2,\y+1)$) circle (0.1);
    }
    \foreach \i in {1,...,5} {
    \pgfmathsetmacro{\x}{rnd}
    \pgfmathsetmacro{\y}{rnd}
    \draw[fill=gray!50] ($(\x,\y+1)$) circle (0.1);
    }
    \foreach \i in {1,...,5} {
        \pgfmathsetmacro{\x}{rnd}
        \pgfmathsetmacro{\y}{rnd}
        \draw[fill=gray!50] ($(\x+1,\y+2)$) circle (0.1);
    }
    \foreach \i in {1,...,5} {
    \pgfmathsetmacro{\x}{rnd}
    \pgfmathsetmacro{\y}{rnd}
    \draw[fill=gray!50] ($(\x+2,\y+2)$) circle (0.1);
    }
    \foreach \i in {1,...,5} {
    \pgfmathsetmacro{\x}{rnd}
    \pgfmathsetmacro{\y}{rnd}
    \draw[fill=gray!50] ($(\x,\y+2)$) circle (0.1);
    }
    \draw (1.5,3.4)node[below]{\textit{Case 1}};
\end{tikzpicture}
\hfill
\begin{tikzpicture}
    \foreach \i in {1,...,5} {
    \pgfmathsetmacro{\x}{rnd*0.4}
    \pgfmathsetmacro{\y}{rnd*0.4}
    \draw[fill=gray!50] ($(\x,\y)$) circle (0.1);
    }
    \foreach \i in {1,...,5} {
    \pgfmathsetmacro{\x}{rnd*0.3}
    \pgfmathsetmacro{\y}{rnd*0.3}
    \draw[fill=gray!50] ($(\x+1,\y)$) circle (0.1);
    }
    \foreach \i in {1,...,5} {
    \pgfmathsetmacro{\x}{rnd*0.3}
    \pgfmathsetmacro{\y}{rnd*0.3}
    \draw[fill=gray!50] ($(\x+2,\y)$) circle (0.1);
    }
    \foreach \i in {1,...,5} {
        \pgfmathsetmacro{\x}{rnd*0.5}
        \pgfmathsetmacro{\y}{rnd*0.5}
        \draw[fill=gray!50] ($(\x+0.5,\y+1)$) circle (0.1);
    }
    \foreach \i in {1,...,5} {
        \pgfmathsetmacro{\x}{rnd*0.4}
        \pgfmathsetmacro{\y}{rnd*0.4}
        \draw[fill=gray!50] ($(\x+2.5,\y+1)$) circle (0.1);
    }
    \foreach \i in {1,...,5} {
        \pgfmathsetmacro{\x}{rnd*0.4}
        \pgfmathsetmacro{\y}{rnd*0.4}
        \draw[fill=gray!50] ($(\x+1.5,\y+1)$) circle (0.1);
        }
    \foreach \i in {1,...,5} {
        \pgfmathsetmacro{\x}{rnd*0.5}
        \pgfmathsetmacro{\y}{rnd*0.5}
        \draw[fill=gray!50] ($(\x,\y+2)$) circle (0.1);
    }
    \foreach \i in {1,...,5} {
        \pgfmathsetmacro{\x}{rnd*0.4}
        \pgfmathsetmacro{\y}{rnd*0.4}
        \draw[fill=gray!50] ($(\x+2,\y+2)$) circle (0.1);
    }
    \foreach \i in {1,...,5} {
        \pgfmathsetmacro{\x}{rnd*0.4}
        \pgfmathsetmacro{\y}{rnd*0.4}
        \draw[fill=gray!50] ($(\x+1,\y+2)$) circle (0.1);
        }
    \draw (1.5,3.4)node[below]{\textit{Case 2}};
\end{tikzpicture}
\hfill
\begin{tikzpicture}
    \foreach \i in {1,...,5} {
    \pgfmathsetmacro{\x}{rnd*1.5}
    \pgfmathsetmacro{\y}{rnd*0.2}
    \draw[fill=gray!50] ($(\x,\y)$) circle (0.1);
    }
    \foreach \i in {1,...,5} {
    \pgfmathsetmacro{\x}{rnd*1.5}
    \pgfmathsetmacro{\y}{rnd*0.3}
    \draw[fill=gray!50] ($(\x+1,\y)$) circle (0.1);
    }
    \foreach \i in {1,...,5} {
    \pgfmathsetmacro{\x}{rnd*1.5}
    \pgfmathsetmacro{\y}{rnd*0.3}
    \draw[fill=gray!50] ($(\x+2,\y)$) circle (0.1);
    }
    \foreach \i in {1,...,5} {
        \pgfmathsetmacro{\x}{rnd*1.5}
        \pgfmathsetmacro{\y}{rnd*0.2}
        \draw[fill=gray!50] ($(\x+1+0.5,\y+1)$) circle (0.1);
    }
    \foreach \i in {1,...,5} {
    \pgfmathsetmacro{\x}{rnd*1.5}
    \pgfmathsetmacro{\y}{rnd*0.2}
    \draw[fill=gray!50] ($(\x+2+0.5,\y+1)$) circle (0.1);
    }
    \foreach \i in {1,...,5} {
    \pgfmathsetmacro{\x}{rnd*1.5}
    \pgfmathsetmacro{\y}{rnd*0.1}
    \draw[fill=gray!50] ($(\x+0.5,\y+1)$) circle (0.1);
    }
    \foreach \i in {1,...,5} {
        \pgfmathsetmacro{\x}{rnd*1.5}
        \pgfmathsetmacro{\y}{rnd*0.2}
        \draw[fill=gray!50] ($(\x+1,\y+2)$) circle (0.1);
    }
    \foreach \i in {1,...,5} {
    \pgfmathsetmacro{\x}{rnd*1.5}
    \pgfmathsetmacro{\y}{rnd*0.2}
    \draw[fill=gray!50] ($(\x+2,\y+2)$) circle (0.1);
    }
    \foreach \i in {1,...,5} {
    \pgfmathsetmacro{\x}{rnd*1.5}
    \pgfmathsetmacro{\y}{rnd*0.1}
    \draw[fill=gray!50] ($(\x,\y+2)$) circle (0.1);
    }
    \draw (1.5,3.4)node[below right]{\textit{Case 3}};
\end{tikzpicture}
\hfill
\caption{Sketch of the three different particle arrangements identified in this study.
The gray disks represent droplets in a three-dimensional space.
Gravity acts in the vertical direction.
(\textit{Case 1: ``homogeneous''}) A homogeneous arrangement of droplets is called a homogeneous microstructure or just ``homogeneous''.
(\textit{Case 2: ``clusters''}) Isotropic non-homogeneous microstructure where we observe the presence of ``clusters'' meaning that the particles are close to each other on average.
The opposite scenario, where particles are, on average, far from each other, also falls into this category.
(\textit{Case 3: ``layers''}) The non-isotropic non-homogeneous microstructure where we observe the presence of stratified arrangements.
This type of microstructure is referred to as layered microstructure or just ``layers''.
}
\label{fig:scheme_clusters}
\end{figure}
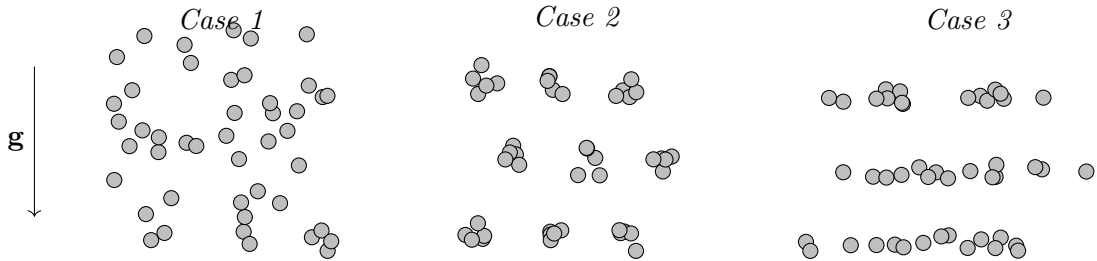

One of the pioneering studies dedicated to the microstructure of bubbly flows via Direct Numerical Simulations (DNS) was undertaken by \citet{bunner2002dynamics}.
They conducted tri-periodic simulations focusing on the behavior of nearly spherical rising bubbles in the regime of moderate inertia, characterized by a bubble Reynolds number of approximately 15.
In their investigation involving around 30 bubbles within a tri-periodic domain, they observed a tendency for the bubbles to align horizontally in pairs, along with the identification of horizontal arrays of particles.
An illustration depicting such a microstructure is provided in \ref{fig:scheme_clusters} (Case 4).
Numerous other studies have investigated the microstructures of bubbly flows using DNS, including those by \citet{yin2008lattice,zhang2021direct}. 
They also observe a tendency towards horizontal alignment of bubble pairs within the microstructure, even at high volume fractions. 

\citet{shajahan2023inertial} investigated the sedimentation of solid spherical particles at high particulate Reynolds numbers.
In the dilute regime, they observed the formation of vertical particle rafts, which transitioned into side-by-side particle arrangements as the particle volume fraction increased. In dense regimes ($\phi \geq 10 \%$), they observed a nearly random distribution of particles. 
\citet{almeras2021statistics} investigated experimentally the microstructure of solid spherical particles within fluidized beds. They noticed the emergence of horizontal particle rafts for moderate volume fraction $\phi \approx 0.2$, which tended to disappear as the volume fraction was increased.  
The studies above solely focused on spherical particles.
It is also known that the deformability of bubbles or more generally, the shapes of the particles greatly influence the microstructure geometry \citep{bunner2003effect,seyed2021sedimentation}.

All these studies reported the formation of particle structures that are dependent on the dimensionless parameters of the problem.
These structures can be classified into three categories: The homogeneous microstructure (\ref{fig:scheme_clusters} (\textit{Case 1: ``homogeneous''})); the non-homogeneous but isotropic microstructure \ref{fig:scheme_clusters} (\textit{Case 2: ``clusters'' }); and the non-homogeneous and non-isotropic microstructure \ref{fig:scheme_clusters} (\textit{Case 3: ``layers''}).
Since the microstructures exhibit a varied range of geometries, having a suitable metric to quantify the microstructure becomes essential. 
Although traditional particle-pair distributions have been employed in previous studies \citep{yin2007,cartellier2009induced, seyed2021sedimentation}, this method has limitations, especially in closure law modeling, as it remains non-zero even at large inter-particle distances. 
Hence, in our study, we adopt the methodology proposed by \citep{zhang2023evolution} and characterize the microstructure using the nearest particle pair distribution.
Specifically, we use a second-order tensor defined as the second moment of nearest neighbor distribution, it is shown to provide a quantitative description of the microstructure.

If one wishes to build relevant statistics, one needs to conduct simulations of droplet flows for sufficient time to reach a statistically steady-state regime.
However, depending on the numerical method used, one may encounter difficulties in preventing coalescence between droplets which is necessary to reach a steady state regime.
Indeed, the Volume-of-Fluid (VoF) method, which is the method used in this work, is known to cause premature coalescence between droplets or bubbles \citep{innocenti2020direct}.
Several methods have been proposed to address this issue, including those by \citet{roghair2011drag,balcazar2015multiple,hidman2023assessing,zhang2023evolution}, and \citet{karnakov2022computing}.
These methods typically involve front-tracking, using a single volume-of-fluid tracer per droplet, or incorporating a numerical interfacial force to hinder coalescence. However, as detailed in the subsequent sections, these methods are only partially suitable for our objectives due to the potential introduction of non-physical behavior or excessive computational overhead. Consequently, we propose a novel algorithm to prevent coalescence, facilitating Direct Numerical Simulation (DNS) over extended periods while maintaining a constant droplet population within a VoF framework.

In this study, we employ tri-periodic Direct Numerical Simulations (DNS) to investigate buoyancy-driven suspensions of mono-disperse droplets. We start by detailing the simulation methodology in \ref{sec:methodo}, presenting our novel algorithm designed to prevent numerical coalescence among droplets. Subsequently, we introduce the Nearest Particle Statistics framework proposed by \citet{zhang2023evolution} in \ref{sec:nearest}. \ref{sec:microstructure} is dedicated to an in-depth examination of the microstructure geometry, where we identify and explore the structures illustrated in \ref{fig:scheme_clusters}, along with their occurrence depending on the dimensionless parameters of the system. Finally, \ref{sec:conclusion} concisely summarizes the key findings presented in this work.

%% file: microstructure/Problem_statement.tex
This section outlines the approach employed for performing simulations to achieve statistically steady states simulations in the context of a rising mono-disperse suspension of droplets within a fully periodic domain.
We start by presenting the relevant physical parameters, followed by an overview of the numerical methods employed.
Finally, we detail the methodology implemented for collecting statistical data on microstructure, which will be presented in the following sections.



\subsection{Problem statement}

We investigate numerically the dynamics of homogeneous mono-disperse emulsions subject to buoyancy forces in a fully periodic domain. 
The dispersed and continuous phases are considered Newtonian fluids defined by viscosity $\mu_d$ (resp. $\mu_f$) and density $\rho_d$ (resp. $\mu_f$).
Throughout this work, the subscript $_d$ and $_f$ indicate properties belonging to the dispersed and continuous phases, respectively. 
The interface between both fluids is considered infinitely thin, free of impurities, and characterized by a constant surface tension $\gamma$. 
The density and viscosity will be considered constant in each phase.
In dimensionless form, this problem is completely characterized by six dimensionless parameters:  the viscosity and density ratio, $\lambda = \mu_d / \mu_f$ and $\zeta = \rho_d / \rho_f$,  
the \textit{Galileo} number, 
\begin{equation*}
    Ga =\frac{\sqrt{\rho_f(\rho_f - \rho_d) g d^3}}{\mu_f},
\end{equation*}
the \textit{Bond} number, 
\begin{equation*}
    Bo =\frac{(\rho_f - \rho_d) g d^2}{\gamma},
\end{equation*}
the number of droplets per domain $N_b$, and the dispersed phase volume fraction $\phi$. 
Here, $d$ represents the diameter of a sphere with the same volume as the droplets and $g$ denotes the acceleration of gravity.
The \textit{Galileo} number measures the strength of the buoyancy forces relative to the viscous forces, whereas the \textit{Bond} number evaluates the ratio between buoyancy and capillary forces. 

In most liquid-liquid systems encountered in industrial processes, the droplet diameters typically range from 10 micrometers to a few millimeters. To illustrate the order of magnitude of the relevant quantities, consider a scenario where vegetable oil is dispersed in water. The continuous phase (water) has a density of approximately $\rho_f = 1000 \text{kg/m}^3$ and a viscosity of about $\mu_f = 10^{-3} \text{Pa.s}$. In contrast, the dispersed phase (vegetable oil) has a density close to $\rho_d = 900 \text{kg/m}^3$ and a viscosity around $\mu_d = 10^{-2} \text{Pa.s}$.
The surface tension of the oil/water system is approximately $\gamma = 0.05 \text{N.m}^{-1}$. The maximum allowable volume fraction is set at $\phi = 0.2$. Beyond this value, particles tend to coalesce easily, leading to a loss of the dispersed flow topology.
\begin{table}[h!]
    \centering
    \caption{Dimensionless parameters of a water/oil system.}
    \begin{tabular}{|c||c|c|c|c|c|}
        \hline&$Ga$&$Bo$&$\phi$&$\lambda$&$\zeta$\\ \hline
        \hline Oil/Water&$[0.35,160]$&$[10^{-5};10^{-1}]$&$<0.2$&$10$&$0.9$\\ \hline
    \end{tabular}
    \label{tab:parameters_exp}
\end{table}
\ref{tab:parameters_exp} gives the corresponding dimensionless parameters.  
Notice that the \textit{Bond number} is relatively low, indicating that the droplets are nearly spherical in these processes.
Following \ref{tab:parameters_exp}, to approach real-life applications, we conducted DNS for four volume fractions, specifically $\phi = 0.01,0.05,0.1,0.2$.
In contrast to most previous studies, we keep the number of droplets constant while changing the volume fraction $\phi$. 
We then modify the domain size $\mathcal{L}$ accordingly. 
This introduces another dimensionless parameter of interest: $\mathcal{L}/d$, which measures the confinement of the particles within the finite numerical domain. 
This parameter is purely determined by $\phi$ and $N_b$, and will thus be refereed as a \textit{secondary parameter}.

As mentioned, the \textit{Bond} numbers of our targeted application is very low.
Therefore, the \textit{Bond} number is set to $Bo = 0.2$, and it will stay constant throughout this study.
DNS with lower \textit{Bond} numbers become excessively expensive due to the restrictive capillary timestep constraint. 
Additionally, the ratio between inertia and surface tension forces is given by the \textit{Weber} number, 
\begin{equation*}
    We = \frac{\rho U^2d}{\gamma}
\end{equation*}
where $U$ is the relative velocity which is the difference between the dispersed phase velocity and the continuous velocity. 
Extreme values of $We$ reached in these simulations are displayed in \ref{tab:simulations}. 
It is clear that for $We=0.6$, we might expect some deformations; nevertheless, in most cases, $We$ stays below these values. 
Consequently, whether in the viscous or inertial regimes, the droplets are expected to remain spherical according to the values of $Bo$ and $We$.
This statement is verified in appendix \ref{ap:deformation}.


The study's primary objective is to investigate the microstructure through the nearest particle pair distribution function.
Thus, obtaining a sufficient number of DNS samples is crucial to ensure statistical convergence. 
Also, the physical quantities measured in the simulations must remain independent of the domain size. 
Therefore, we use a number of particles per domain of $N_b = 125$, roughly what \citet{hidman2023assessing} used for their DNS of fully-periodic buoyant rising bubbles.
Moreover, each DNS lasts for a time: $t^*_\text{end} = 1500 \sqrt{d/g}$.
\begin{table}[h!]
    \centering
    \caption{Dimensionless parameter range investigated in this work.}
    \begin{tabular}{|ccccccc|ccc|}\hline
        \multicolumn{7}{|c|}{Primary parameters}&\multicolumn{3}{|c|}{Secondary parameters}\\\hline\hline
        $Ga$&$Bo$&$\phi$&$\lambda$&$\zeta$&$N_b$&$t^*_\text{end}$&$\mathcal{L}/d$&$Re$&$We$\\ \hline
        $5\rightarrow 100$&$0.2$&$1\% \rightarrow 20\%$&$10$ \& $1$&$0.9$&$125$&$1500$&$6.7\to 18.7$&$10^{-1}\to 170$&$10^{-4}\to 0.6$\\ \hline
    \end{tabular}
    \label{tab:simulations}
\end{table}
This study presents DNS results with dimensionless parameters in ranges outlined in \ref{tab:simulations}.
In summary, we investigated $5$ \textit{Galileo} number $Ga = 5,10,25,50,100$, $4$ different volume fractions $\phi = 0.01,0.05,0.1,0.2$, and two viscosity ratios $\lambda =1,10$ with $Bo = 0.2$ and $\zeta = 0.9$. 
This makes a total of $40$ representative simulations of $N_b = 125$ droplets which last for $t= 1500 \sqrt{d/g}$. 

%% file: microstructure/Simulation_Setup.tex
\subsection{Numerical method}
The governing equations describe the motion of two immiscible fluids of different densities and viscosities separated by an interface with surface tension. 
We use the so-called ``one fluid'' formulation of the variable density and viscosity Navier-Stokes equations, which can be expressed as \citep{tryggvason2011direct}
\begin{align}
    \pddt \rho+ \div(\rho\textbf{u})
    &= 0,\\
    \label{eq:dt_urho}
    \pddt (\rho \textbf{u})
    + \div (\rho  \textbf{u} \textbf{u} - \bm\sigma)
    &= (\avg{\rho} - \rho)\textbf{g}
    + \textbf{f}_\gamma,\\
    \label{eq:dt_C}
    \pddt C + \textbf{u}\cdot\grad C  
    &= 0,
\end{align}
for the mass, momentum and colors function transport equations, respectively. 
The scalar field, $C$, represents the color function, which ranges between $0$ and $1$ to indicate the volumetric proportion of each phase.
We introduced the fluid velocity vector $\textbf{u}$ and the Newtonian stress tensor $\bm{\sigma} = -p \bm\delta + \mu (\grad \textbf{u}+ \grad \textbf{u}^\dagger)$ where $p$ is the pressure field, $^\dagger$ represents the transpose operator and $\bm\delta$ the unit tensor.
Note that the material properties, $\rho$, and $\mu$, take the values of each phase in presence, using the arithmetic average: $\rho = (1-C)\rho_f + C \rho_d$ and $\mu = (1-C)\mu_f + C \mu_d$. 
In our case, the arithmetic mean performs better than the harmonic mean, which is often used to interpolate the viscosity for bubbly flows \citet{hidman2023assessing,innocenti2020direct}.
More details about this choice are provided in \ref{ap:validation} (\textit{Case 1.}). 
The capillary force is defined as $\textbf{f}_\gamma =\textbf{n} \gamma \div \textbf{n} $, where \textbf{n} is the normal to the interface.
Following  \citep{bunner2002dynamics}, we incorporated the artificial body force term, $\avg{\rho}\textbf{g}$, on the right-hand side of \ref{eq:dt_urho}, to mimic a zero-averaged velocity throughout the entire numerical domain.  

Before solving these equations, we first initialized $125$ spherical droplets within a cubic domain with fully periodic boundary conditions. 
We used the open source code \url{http///basilisk.fr} to discretize the governing equations. 
The Navier-Stokes equations are discretized with a centered scheme.
The two-phase flow solver uses the geometric Volume of Fluid (VoF) method. 
The interfaces between the droplets and the carrier fluid is reconstructed using the Piecewise Linear Inter-face Calculation or PLIC method \citet[Chapter 5.]{tryggvason2011direct}.
Regarding treating the surface tension force term, we refer the reader to \citet{popinet2018numerical} for more details. 
The Basilisk solver has been validated extensively in the framework of bubbly flows. 
Most previous studies \citep{hidman2023assessing,innocenti2020direct} recommend a resolution of $\Delta/d \ge  30$, where $\Delta$ is the grid spacing. 
In \ref{ap:validation}, we carry out a mesh-independence study and demonstrate that a grid spacing of $\Delta/d = 20$ is suitable in our context.
For readers seeking more detailed information about the solvers, we recommend the wiki pages: \href{http://basilisk.fr/src/navier-stokes/centered.h}{centered.h}, \href{http://basilisk.fr/src/tension.h}{tension.h} and \href{http://basilisk.fr/src/poissson.h}{poissson.h} where one can find the source code of the Navier--Stokes, surface tension and multigrid solver used in this work, respectively. 

With the VoF method, droplets and bubbles may experience premature coalescence.
See \citet[Appendix B]{innocenti2020direct} for a detailed discussion on this issue.
However, in this work, it is imperative to conserve a specific (mono-disperse) population of droplets over time to accumulate sufficient statistics about the microstructure.
To tackle this issue, we present in the next section a novel algorithm that prevents coalescence between droplets at a reasonable computational cost. 
Note that the wiki page \href{http://basilisk.fr/sandbox/fintzin/Rising-suspension/RS.c}{RS.c} complements this section, where the reader can access the source code used to conduct these DNS, as well as comments and notes to help comprehension.

%% file: microstructure/no_coalesence.tex
\subsection{The \texttt{no-coalescence.h} algorithm}

In previous studies, various methods have been used to avoid coalescence. 
One method is to artificially increase the surface tension coefficient at the interface contact points, as demonstrated in the recent study of \citet{hidman2023assessing}.
However, it remains unclear if the physical behavior of the droplets interactions is well captured due to the introduction of artificial forces. 
Additionally, its applicability for denser emulsions, up to $\phi = 0.2$, remains to be determined. 
\citet{balcazar2015multiple} developed a multiple-marker level-set method to prevent coalescence, while \citet{zhang2021direct} used a multi-VoF method. 
The latter method involves assigning each droplet a different color function so that the interfaces are reconstructed independently when the droplets are in close contact.
Since the representations of the interfaces are independent, droplets in contact never coalesce.  
The latter method may be suitable for our objectives; however, it can be quite expensive as it requires solving a transport equation for each tracer, with one tracer assigned per droplet, which means $125$ tracers in our case. 
\citet{karnakov2022computing} developed a multi-VoF method that requires a fixed number of tracers for an arbitrary number of droplets.
This approach allows multiple non-touching droplets to belong to the same field, which makes it more efficient than the previous study.
Although this approach shares the same basic principle as the one used here, i.e., coloring adjacent droplets with different colors to avoid coalescence, it differs in terms of computational methods. 
In the following, we present our methodology that is implemented in the \texttt{Basilisk} framework. 

The challenge here is to assign a tracer to every adjacent droplet to prevent numerical coalescence while minimizing the number of tracers to reduce computational cost. 
This recalls the famous \textit{Four color map theorem} \citep{appel1977solution} which essentially states that : 
\enquote{every map can be color using only four colors, so that two neighboring region are different colors}. 
In our case, this theorem implies that for any 2D configuration only four VoF tracer are necessary to avoid coalescence\footnote{It is worth noting that for a bi-periodic domain, seven colors are required due to the torus-like topology.  }. 
Therefore, leveraging the \textit{Four color map theorem}, one might be able to significantly reduce the number of VoF tracers required.
Note however that the optimal coloring problem is originally a static problem that need to be solved only once. 
In our case, droplets move around over time thus transforming the static problem into a time-dependent problem. 
Furthermore finding the optimal coloring is known to be difficult (the problem is {\em NP-complete}) and solving it at each timestep would be too expensive.

Note also that in three-dimensions, the \textit{Four color map theorem} has no equivalent.
For example, an arbitrarily large number of rectangular blocks in 3D space can all touch each other, requiring an arbitrarily large number of colors to differentiate the adjacent blocks \citep{magnant2011coloring}. 
We are not aware of an extension of the coloring problem to arrangements of spheres in 3D. 
Nevertheless, it is reasonable to assume that the number of tracers required to avoid coalescence is significantly smaller than the number of droplets.
Consequently, since we cannot determine the optimal coloring configuration based on theoretical grounds, we assign the tracers to each droplet following the empirical strategy detailed below.

The development of the \texttt{no-coalescence.h} algorithm was initiated in the PhD. thesis of \citet{mani2021numerical}.
The latest version of this algorithm can be found on the Basilisk wiki page: \href{http://basilisk.fr/src/no-coalescence.h}{no-coalescence.h}.
Before diving into a step-by-step description of this algorithm, we must introduce another key feature used in these simulations: the \href{http://basilisk.fr/src/tag.h}{tag.h} algorithm. 
It is an adaptation of the \textit{painter}'s algorithm but optimized using the multigrid solver of \texttt{Basilisk}. 
Its purpose is to assign different scalar values to each cell belonging to different regions, with the regions being delimited by the different droplets' interfaces. 
For instance, on \ref{fig:images} (left), we can see two blue regions corresponding to two different drops.
We notice that both are assigned with two different values, $1$ and $2$, which are identified using the \texttt{tag.h} algorithm. 
It is then straightforward to obtain the droplet properties, such as its center of mass, by carrying numerical integration on the VoF field, considering only the cells having a specific tag value that corresponds to a given droplet.  

\begin{figure}
    \centering
    \begin{tikzpicture}[scale=0.5,ultra thick]
        \def\nRows{10}
        \def\nCols{20}
        \pgfmathsetmacro\nRowsm{\nRows-1}
        \pgfmathsetmacro\nColsm{\nCols-1}

        \foreach \row in {0,...,\nRowsm} {
            \foreach \col in {0,...,\nColsm} {
                \pgfmathsetmacro\distance{veclen(\col-4.356, \row-2.65)};
                \pgfmathparse{\distance < 4 ? "blue" : "white"}
                \edef\colour{\pgfmathresult};
                \ifthenelse{\equal{\colour}{blue}}{                    
                    \fill[\colour!60!white] (\col, \row) rectangle ++(1,1);
                    \node (num) at (\col +0.5,\row+0.5){1};
                }
            }
        }

        \foreach \row in {0,...,\nRowsm} {
            \foreach \col in {0,...,\nColsm} {
                \pgfmathsetmacro\distance{veclen(\col-15, \row-6.2)};
                \pgfmathparse{\distance < 3.5 ? "blue" :"white"}
                \edef\colour{\pgfmathresult};
                \ifthenelse{\equal{\colour}{blue}}{
                    \fill[\colour!60!white] (\col, \row) rectangle ++(1,1);
                \node (num) at (\col +0.5,\row+0.5){2};
                }
            }
        }

        \foreach \row in {0,...,\nRowsm} {
            \foreach \col in {0,...,\nColsm} {
                \pgfmathsetmacro\distance{veclen(\col-15.62, \row-1.5)};
                \pgfmathparse{\distance < 2.5 ? "yellow" :"white"}
                \edef\colour{\pgfmathresult};
                \ifthenelse{\equal{\colour}{yellow}}{
                    \fill[\colour] (\col, \row) rectangle ++(1,1);
                    \node (num) at (\col +0.5,\row+0.5){2};
                }
            }
        }
        \pgfmathsetmacro\gridSize{1}
        
        \foreach \row in {0,...,\nRows} {
            \draw [gray!50] (0,\row*\gridSize) -- (\nCols*\gridSize,\row*\gridSize);
        }
        \foreach \col in {0,...,\nCols} {
            \draw [gray!50] (\col*\gridSize,0) -- (\col*\gridSize,\nRows*\gridSize);
        }
        \draw[red] (8,2)node[below right]{stencil} rectangle +(5,5); 
        \filldraw[gray] (10,4) rectangle +(1,1);
    \end{tikzpicture}    
    \begin{tikzpicture}[scale=0.5,ultra thick]
        \def\nRows{10}
        \def\nCols{10}
        \pgfmathsetmacro\nRowsm{\nRows-1}
        \pgfmathsetmacro\nColsm{\nCols-1}

        \foreach \row in {0,...,\nRowsm} {
            \foreach \col in {0,...,\nColsm} {
                \pgfmathsetmacro\distance{veclen(\col, \row-2)};
                \pgfmathparse{\distance < 3.5 ? "blue" :"white"}
                \edef\colour{\pgfmathresult};
                \ifthenelse{\equal{\colour}{blue}}{
                    \fill[\colour!60!white] (\col, \row) rectangle ++(1,1);
                \node (num) at (\col +0.5,\row+0.5){1};
                }
            }
        }

        \foreach \row in {0,...,\nRowsm} {
            \foreach \col in {0,...,\nColsm} {
                \pgfmathsetmacro\distance{veclen(\col-7, \row-2)};
                \pgfmathparse{\distance < 3.5 ? "blue" :"white"}
                \edef\colour{\pgfmathresult};
                \ifthenelse{\equal{\colour}{blue}}{
                    \fill[\colour!60!white] (\col, \row) rectangle ++(1,1);
                    \node (num) at (\col +0.5,\row+0.5){1};
                }
            }
        }
        \foreach \row in {0,...,\nRowsm} {
            \foreach \col in {0,...,\nColsm} {
                \pgfmathsetmacro\distance{veclen(\col-5, \row)};
                \pgfmathparse{\distance < 3.5 ? "blue" :"white"}
                \edef\colour{\pgfmathresult};
                \ifthenelse{\equal{\colour}{blue}}{
                    \fill[\colour!60!white] (\col, \row) rectangle ++(1,1);
                    \node (num) at (\col +0.5,\row+0.5){1};
                }
            }
        }
        \foreach \row in {0,...,\nRowsm} {
            \foreach \col in {0,...,\nColsm} {
                \pgfmathsetmacro\distance{veclen(\col, \row-9)};
                \pgfmathparse{\distance < 3.5 ? "blue" :"white"}
                \edef\colour{\pgfmathresult};
                \ifthenelse{\equal{\colour}{blue}}{
                    \fill[\colour!60!white] (\col, \row) rectangle ++(1,1);
                    \node (num) at (\col +0.5,\row+0.5){1};
                }
            }
        }
        \pgfmathsetmacro\gridSize{1}
        
        \foreach \row in {0,...,\nRows} {
            \draw [gray!50] (0,\row*\gridSize) -- (\nCols*\gridSize,\row*\gridSize);
        }
        \foreach \col in {0,...,\nCols} {
            \draw [gray!50] (\col*\gridSize,0) -- (\col*\gridSize,\nRows*\gridSize);
        }
        \draw[red] (2,3) rectangle +(5,5)node[below left]{stencil}; 
        \filldraw[gray] (4,5) rectangle +(1,1);
    \end{tikzpicture}    
    \caption{Sketch of two situations where the \textit{near contact} criterion is true. 
    The background grid represents the cells within the numerical domain. 
    The dark blue area represents the cells where $C_i > 0$.
    The yellow area represents the cells where the tracer $C_j > 0$ for $j\neq i$. 
    The numbers represent the values of the $Tag$ scalar field within each tracer.
    The $5$ by $5$ cells red rectangle represents the stencil zone, which iterates over all domain cells.  
    (left) Two droplets are in contact since we have two opposite cells in the stencil with $C_i > 0$ and $C_i=0$ at the center.
    Moreover, the mentioned cells belong to two different regions, so (\textit{Step 3}) is also validated.  
    (right) A near contact is observed since we have two opposite cells in the stencil with $C_i > 0$ and $C_i=0$ at the center however in this case, we do not verify the second criterion of \textit{Step 3}, which requires two different tags values. 
    }
    \label{fig:criterion}
\end{figure}
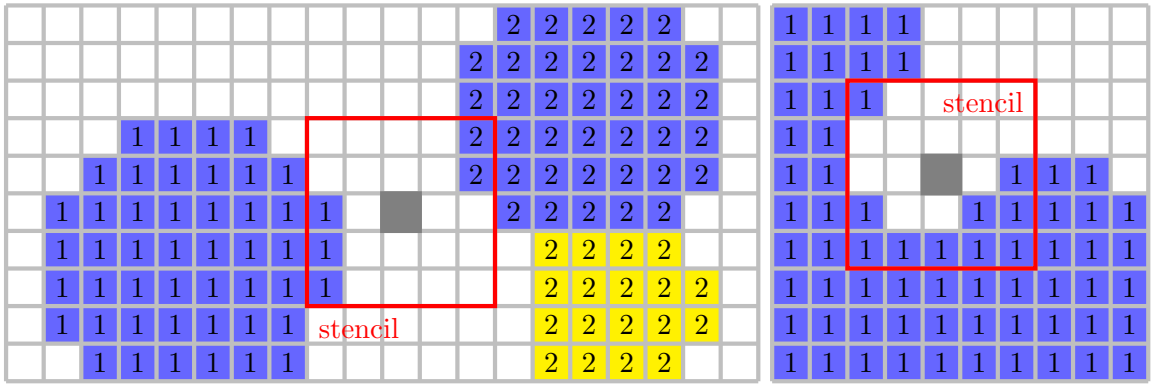

\begin{figure}[h!]
    \centering
    \begin{tikzpicture}
    \node (img) at (0,0) {\includegraphics[width = 0.4\textwidth]{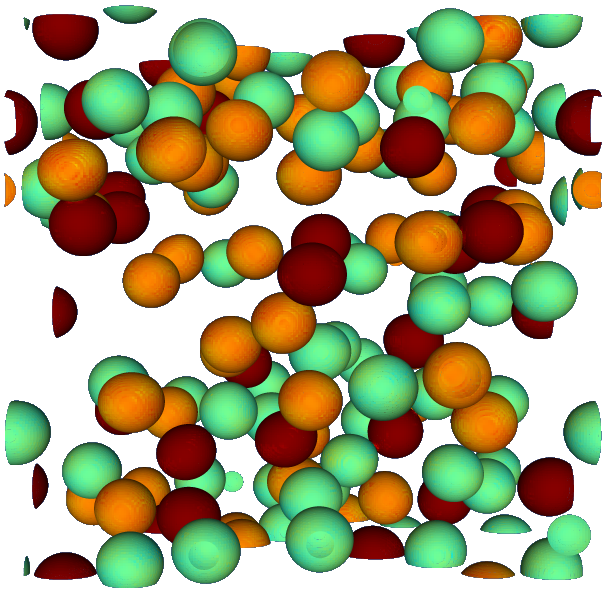}};
    \node (img) at (0.4\textwidth,-0.01\textwidth) {\includegraphics[width = 0.4\textwidth]{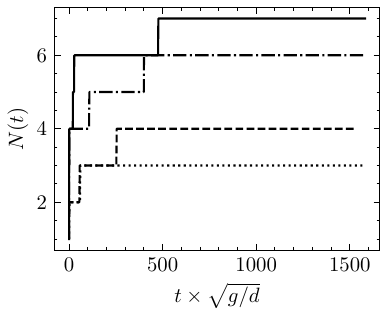}};
    \end{tikzpicture}
    \caption{
    (left) Snapshot of a DNS with $\phi = 0.05$, $\lambda = 1$, $Ga = 50$ with the interface of the droplets colored by the index of the tracers.
    (right) Number of tracers $N(t)$ as a function the dimensionless time.
    Four different volume fractions are displayed : (dotted line) $\phi = 0.01$, (dashed line) $\phi = 0.05$ (dash dotted line) $\phi = 0.1$ (solid line) $\phi = 0.2$ at $Ga = 50$ and $\lambda = 1$. 
    }
    \label{fig:diagram}
\end{figure}

We define the $i^\text{th}$ color function as $C_i$ for $i =1,2,\ldots,N(t)$, where $N(t)$ is the total number of tracers used in a simulation at time $t$.
The color function introduced previously is now defined as, $C = \sum_{i=1}^{N(t)} C_i$. 
Note that $N(t)$ is time-dependent since the number of tracers may increase during the simulation as droplets get closer to one another.
The simplified workflow of the algorithm follows these four steps : 
\begin{enumerate}
    \item[\textit{Step 1}.] Check if, within a tracer field $C_i$, the droplets are possibly too close to each other. 
    The \textit{near contact} criterion that determines if the droplets are too close is defined using a $5$ by $5$ cells stencil, which verifies the following conditions : 
    (1) If the color function $C_i = 0$ at the center of the stencil. 
    (2) And if $C_i > 1$ for two opposite cells in the stencil. 
    In this case, two different regions might be in close contact.
    A sketch of this situation is given in \ref{fig:criterion}.  
    \item[\textit{Step 2}.] 
    If (\textit{Step 1}.) is true for the tracer $C_i$, we must verify if we indeed identified two different regions in near contact and not just a single region close to itself, such as in \ref{fig:diagram} (right). 
    Therefore, at this step, we apply the \texttt{tag.h} algorithm.
    \item[\textit{Step 3}.] Re-use the \textit{near contact} criterion of (\textit{Step 1}.) by requiring in addition that the cells must belong to two different tag groups. 
    At this stage, the situation in \ref{fig:criterion} (left) would be true, while the situation on \ref{fig:criterion} (right) would be false. 
    We, therefore, identified all the droplets / regions that are indeed too close to each other. 
    \item[\textit{Step 4}.] 
    Find a new tracer field $C_n$, with which we could set the region/droplets that are in near contact. 
    At this stage, it is essential to identify the list of tracers $C_j$ already in contact with the region to be replaced. 
    For example, in \ref{fig:criterion} (left), the droplet on the right is adjacent to a region with tracer $C_j$, in which case $C_n$ must satisfy $n \neq i,j$. 
    Therefore, any $n$ in $1, 2, \ldots, N(t)$ are suitable candidates if this situation is avoided. 
    If the droplet is already adjacent to every tracer $C_j$ in the simulation, for $j = 1, 2, \ldots, N(t)$, we create a new tracer $C_n$ with $n = N(t)+1$ and assign the drop to this tracer field.
\end{enumerate}
These four steps are executed at each simulation time step and for each tracer $C_i$ with $i = 1, 2, \ldots, N(t)$.
Following this procedure, we ensure that all adjacent droplets use different tracers, ultimately preventing coalescence. 

Having $N(t)$ tracers requires modifications to the aforementioned governing equations. 
Specifically, instead of solving \ref{eq:dt_C}, we solve $N(t)$ transport equations, one for each $C_i$.
Likewise, the surface tension force is computed as the sum of the contributions from each $C_i$ and reads
\begin{align*}
    \pddt C_i + \textbf{u}\cdot\grad C_i = 0,
    \ \  \ \ \forall i = 1,2,\ldots N(t),\\
    \textbf{f}_\gamma 
    = \sum_{i=0}^{N(t)} \gamma \kappa_i \grad C_i
\end{align*}
where $\kappa_i$ is the numerical approximation of the curvature of the interface of field $C_i$, which is computed following the same method employed for a single tracer. 

\ref{fig:diagram} (left) shows a snapshot of a simulation at an arbitrary time $t^* = 100 \sqrt{g/d}$. 
The droplet interfaces are colored by the index of their respective tracer. 
In this simulation at that time, no more than 3 colors are needed to avoid coalescence.
On \ref{fig:diagram} (right), we display the value of $N(t)$ in terms of the dimensionless simulation time for various volume fractions $\phi$ at $Ga = 50$ and  $\lambda = 1$. 
We observe that for the entire simulation, no more than three tracers were needed for the dilute emulsion ($\phi = 0.01$) and up to 7 for $\phi = 0.2$. 
Although our algorithm could be optimized, it brings sufficient efficiency for our needs. 
In particular, it is observed that the \texttt{no-coalesce.h} algorithm accounts for approximately $4\%$ of the total computational time of a simulation in the densest scenario. As for the \texttt{tag.h} algorithm, its cost is around $2\%$, which is also reasonable. In comparison, the \texttt{poisson.h} solver is about $13\%$ of the simulation time. 
To reduce the cost related to the \texttt{no-coalesce.h} algorithm we believe that further developments, which are referenced at \href{http://basilisk.fr/src/no-coalescence.h}{no-coalescence.h}, are still doable and could be useful for future studies.

In order to validate our numerical methodology, we have compared our numerical results to the experiments of \citet{mohamed2003drop}, where they experimentally study the impact of a single drop on a flat interface of the same fluid. 
It is found that the multi-VoF method captures the position of interfaces remarkably well, even with a poor description of the liquid film between these interfaces.
Additionally, we argue that the mesh independence study conducted in \ref{ap:validation} (Case 3) substantiates the accuracy of the DNS, as the dynamics of interaction converge with reasonable errors for a grid resolution of $\Delta/d = 20$. 
Overall, we used an optimized multi-VoF method, enabling us to perform DNS with a maximum of 7 tracers in the densest scenario.

%% file: microstructure/The_nearest_particle_statistics.tex
We adopted the \textit{nearest particle statistics} framework recently revisited by \citet{zhang2021ensemble} to study the emulsion microstructure.
We now recall some definitions of the \textit{nearest particle statistics} averaging procedure. 
For further details, readers are encouraged to refer to \citet{zhang2021ensemble,zhang2023evolution} on which this work mainly relies.

Let $P(\FF)$ be the probability density function that describes the probability of finding the flow in the configuration $\FF$, where $\FF = (\lambda_1,\lambda_2,\lambda_3,\ldots)$ is the set of all parameters describing the flow configuration.
We then define $d\mathscr{P} = P(\FF)d\FF$ as the probable number of realizations in the incremental region of the flow phase space, $d\FF$ around $\FF$.
Additionally,  $\textbf{x}_i(t,\FF)$ and $\textbf{x}_j(\FF,t)$ refer to the Lagrangian position vectors of the particles $i$ and $j$, respectively. 
Note that the particle positions are functions of time and of the flow configuration $\FF$. 
The nearest pair probability density function is given by \citep{zhang2021ensemble,zhang2023evolution}
\begin{equation}
    P_{nst}(\textbf{r}|\textbf{x},t)= \frac{1}{n_p(\textbf{x},t)}
    \int \sum_{i}^{N_b}\delta[\textbf{x}-\textbf{x}_i(t,\FF)]
    \sum_{j\neq i}^{N_b}\delta[\textbf{x}+\textbf{r}-\textbf{x}_j(t,\FF)]
    h_{ij} (t,\FF)
    d\mathscr{P},
    \label{eq:P_nstij}
\end{equation}
where we introduced the function 
\begin{equation*}
    h_{ij}(\FF,t)
    = \left\{
        \begin{tabular}{cc}
            $1/N_i(\FF,t)$ & if $j$ is one of the $N^{th}_i$ nearest neighbors of $i$ \\
            0& otherwise
        \end{tabular}
        \right. ,
\end{equation*}
and the number density, 
\begin{equation}
    n_p(\textbf{x},t)= 
    \int \sum_{i}^{N_b}\delta[\textbf{x}-\textbf{x}_i(t,\FF)] d\mathscr{P}.
    \label{eq:n_p}
\end{equation}
$P_{nst}(\textbf{r}|\textbf{x},t)$ is the probability density of finding the nearest neighbor at $\textbf{x}+\textbf{r}$ given a particle already in \textbf{x}.
In this definition, we considered the situation where $N_i(\FF,t)$ particles could be simultaneously nearest neighbors to the particle $i$. 
In the DNS, having two nearest neighbors to one particle may never occur; thus, in most cases, $h_{ij}$ is either $1$ or $0$. 
Nevertheless, the coefficient $1/N_i$ must be retained in the definition for theoretical consistency.
Note that $P_\text{nst}(\textbf{r}|\textbf{x},t)$ is a probability density, therefore we have
\begin{equation*}
    \int_{\mathbb{R}^3}
     P_\text{nst}(\textbf{r}|\textbf{x},t) d\textbf{r}  = 1. 
    \label{eq:Pnst}
\end{equation*}

To collect statistics from the DNS, each simulation timestep is treated as an independent flow configuration. 
Data for each Lagrangian quantity were gathered every 10 simulation timesteps. 
The simulation timestep is determined either by the Courant-Friedrichs-Lewy (CFL) condition or by the capillary timestep, depending on the relevant dimensionless numbers.
On average, $200,000$ timesteps are performed during a simulation with $N_b = 125$ droplets. 
This results in a total of $E = 2,500,000$ events.
$P_\text{nst}(\textbf{r}|\textbf{x},t)$ is obtained by averaging over all events $E$. 
Consequently, for our concern, $P_\text{nst}(\textbf{r}|\textbf{x},t)$ is not dependent on $\mathbf{x}$ and $t$, but remains a function of the global flow parameters :  $Ga$, $\phi$, $Bo$, $\zeta$, and $\lambda$.
Therefore, in the following we drop $\mathbf{x}$ and $t$ in our notation. 

To reconstruct $P_\text{nst}(\textbf{r})$ from DNS all the relative positions $\textbf{r}_{ij}  = \textbf{x}_j - \textbf{x}_i$ are measured and stored in $n$ intervals of the form $[\textbf{r}_k; \textbf{r}_k+\Delta \textbf{r}]$ for $k = 1,\ldots, n$ with $n$ being a positive integer.
Then, the nearest-neighbor probability density function can be obtained discretely as follows
\begin{equation}
    P_\text{nst}(\textbf{r}_k)
    =\lim_{\Delta \textbf{r} \to \bm 0} \frac{E_k}{E \Delta \textbf{r}},
    \label{eq:vec_cond}
\end{equation}
with $E_k$ the total number of events where the nearest particle pair verifies $\textbf{r}_{ij} \in [\textbf{r}_k ; \textbf{r}_k + \Delta \textbf{r}]$.
Note that $P_\text{nst}(\textbf{r}_k)$ takes discrete values as arguments and is equivalent to $P_\text{nst}(\textbf{r})$ only in the limit $\Delta \textbf{r} \to 0$. 
Furthermore, it is convenient to represent the vector \textbf{r} using radial, polar, and azimuthal coordinates,  $r = |\textbf{r}|$,$\beta$ and $\theta$, respectively. 
$\theta$ is the angle between the vector \textbf{r} and the vertical direction and $\beta$ the polar angle defined from $0$ to $2\pi$. 
In particular, due to the symmetries of the problem under consideration, we assume that
\begin{equation}
    P_\text{nst}(r,\theta)
    = P_\text{nst}(r,\theta,\beta). 
\end{equation}
Let us define $[r_k; r_k+\Delta r]$ and $[\theta_k; \theta_k+\Delta \theta]$  as $2k$ intervals of radial and polar coordinates, respectively. 
Then, the probability density $P_\text{nst}(r,\theta)$ is obtained using the following formula
\begin{equation}
    P_\text{nst}(r_k,\theta_k)
    =
    \lim_{\Delta \theta , \Delta r \to 0}
    \frac{E_k}{E}
    \frac{1}{2\pi  r_k^2 \sin \theta_k \Delta r \Delta \theta}.
    \label{eq:Ptheta_r}
\end{equation}
In this case, $E_k$ is the total number of events where the nearest particle pair verifies $r_{ij} \in [r_k ; r_k + \Delta r]$, and $\theta_{ij} \in [\theta_k; \theta_k+d\theta]$, and $2\pi  r_k^2 \sin \theta_k$ serves as the normalization factor. 
Now let us define the radial probability density function  $P_\text{r-nst}(r)$ as the average of $P_\text{nst}(\textbf{r})$ over the surface of a sphere centered at $\textbf{r}=\textbf{0}$ with radius $r$, namely
\begin{equation}
    P_\text{r-nst}(r) = \frac{1}{4\pi }\int_{0}^{2\pi}\int_{0}^{\pi} P_{nst}(r,\theta ,\beta) \sin\theta d\theta d\beta.
    \label{eq:P_r}
\end{equation}
Notice that if $P_\text{nst}$ is independent of $\theta$ and $\beta$, then $P_\text{nst-r} = P_\text{nst}$. 
The definition of $P_\text{nst-r}$ based on simulation samples is given by 
\begin{equation}
    P_\text{r-nst}(r_k)
    =
    \lim_{\Delta r \to 0}
    \frac{E_k}{E}
    \frac{1}{4\pi  r_k^2  \Delta r }
    \label{eq:P_r2}
\end{equation}
where $E_k$ is the total number of events where the nearest particle pair verifies $r_{ij} \in [r_k ; r_k + \Delta r]$, and $4\pi  r_k^2  \Delta r$ corresponds to the volume of the spherical shell containing these neighboring particles. 

\ref{eq:Ptheta_r} and \ref{eq:P_r2} are exact only in the limit $\Delta r,\Delta \theta  \to 0$. 
However, due to limitations in the number of available samples, we must choose small but finite intervals.
In this work, we use $\Delta \theta = \frac{\pi}{150}$ and $\Delta r = \frac{1}{100}d$, as it will be shown to provide satisfactory results. 
For sake of clarity, we omit the distinction between $P_\text{nst}(r_k,\theta_k)$ and $P_\text{nst}(r,\theta)$ in the next sections. 

%% file: microstructure/Non_Inertial.tex
This section presents an analysis of the microstructure based on the nearest neighbor probability density function. 

By definition, $P_\text{nst}(\textbf{r}|\textbf{x},t)$ does not require symmetry with respect to the variable $\textbf{x}+\textbf{r}$ and \textbf{x}, as is the case for classical particle-pair distribution functions. 
Nevertheless, it turns out that $P_\text{nst}$ possesses a nearly-symmetric distribution, such that  $P_\text{nst}(r,\theta)\approx P_\text{nst}(r,- \theta)$ as demontrated below.
\begin{figure}[h!]
    \centering
    \includegraphics[height = 0.3\textwidth]{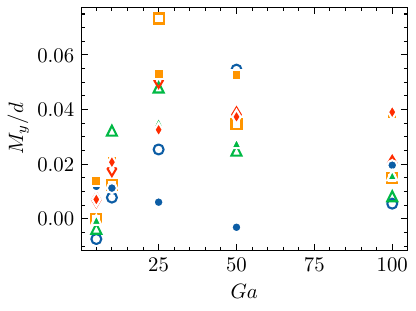}
    \caption{ Dimensionless first moment of the nearest particle pair distribution in the direction of gravity $M_y/d$. 
    ($\pmb\bigcirc$) $\phi = 0.01$; ($\pmb\triangle$) $ \phi = 0.05$; ($\pmb\square$) $\phi = 0.1$ ($\pmb\lozenge$) $\phi = 0.2$.
    The hollow symbols correspond to $\lambda = 1$, the filled symbols to $\lambda = 10$.
    }
    \label{fig:ap:RY}
\end{figure}
We define the first moment of $P_\text{nst}$ as
\begin{equation}
 \textbf{M} = \int_{\mathbb{R}^3} \textbf{r} P_\text{nst}(\textbf{r}) d\textbf{r}.
\end{equation}
\ref{fig:ap:RY} illustrates the projection of $\textbf{M}$ along the direction of gravity. 
The relatively small but finite values of $M_y$ indicate that $P_\text{nst}$ exhibits a nearly symmetric distribution with respect to $\theta$.
Nevertheless, this indicates that the nearest neighbor is more likely to be located in the upstream direction. 
Note that this is consistent with the findings of \citet{zhang2023evolution}.
Even though this slight asymmetry might have its importance \cite{zhang2023evolution}, we discard it in this study. 
Therefore, we choose to show only the upper part of the distribution in the following plots, (displayed in \ref{fig:Pnst_low_Ga} and \ref{fig:Pnst_high_Ga}) since qualitatively it remains the same as the lower part.  
Additionally, in the discussion below, we refer to the sphere at the origin of the graphs, located at $\textbf{x}=0$, as the \textit{test particle}.

\subsection{Low inertia regimes}
We begin with a detailed analysis of $P_\text{nst}$ at $Ga =10$, to investigate the influence of $\lambda$ and $\phi$ on the microstructure when inertial effects are small.
\begin{figure}[h!]
    \centering
    \includegraphics[height=0.21\textwidth]{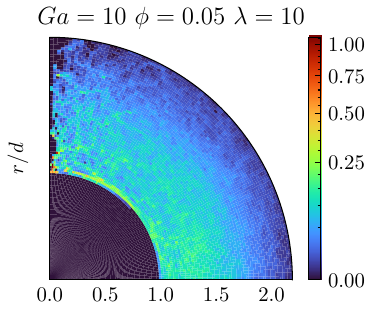}
    \includegraphics[height=0.21\textwidth]{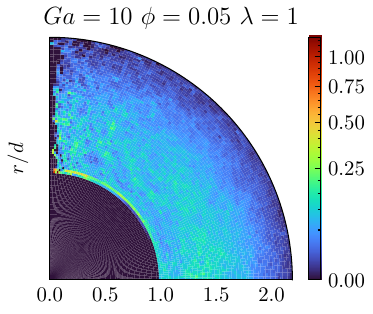}
    \includegraphics[height=0.21\textwidth]{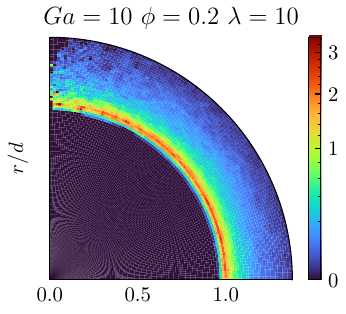}
    \includegraphics[height=0.21\textwidth]{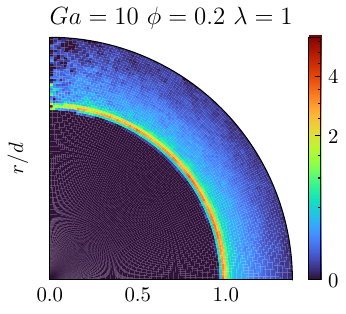}
    \caption{Histogram of the probability density function, $P_\text{nst}(r,\theta)$, for low inertia $Ga = 10$.
    The color map represents the nearest pair distribution function. 
    The origin corresponds to the position of the \textit{test particle}.
    The dimensionless radial and azimuthal coordinates, $|\textbf{r}|/d$ and $\theta$, correspond to the nearest neighbor position.
    The vertical direction corresponds to the flow direction, which is also the axis of symmetry for $P_\text{nst}$.
    (left) Low volume fraction cases $\phi=0.05$ for $\lambda = 1,10$.
    (right) High volume fraction cases $\phi=0.2$ for $\lambda = 1,10$.
    }
    \label{fig:Pnst_low_Ga}
\end{figure}
The first observation from \ref{fig:Pnst_low_Ga} indicates that the likelihood of finding the nearest neighboring particle at an angle $\theta$ is uniform across all $\theta$.
This suggests that $P_\text{nst}$ is isotropic at these \textit{Galileo} numbers. We can observe that $P_\text{nst}$ is larger close to the \textit{test particle} ($r/d = 1$) in the high volume fraction cases than in the low volume fraction cases.
In practice, if particles are more likely to be close to one another, it means that densely packed regions of particles are present in the flow.
This suggests that isotropic clusters, as represented in \ref{fig:scheme_clusters} (\textit{Case 2}), are likely to form in the present context. 
Regarding the effect of the viscosity ratio, $P_\text{nst}$ are very similar for both values of $\lambda$ except that for the highest volume fraction the region of highest probability is thinner for the lowest aspect ratio. 
Consequently, in this regime we find homogeneous microstructures at low $\phi$, and non-homogeneous but still isotropic microstructure (\textit{``clusters''}) at higher $\phi$.

%% file: microstructure/Inertial.tex
\subsection{High inertia regimes }
We now focus on the high inertia regimes ($Ga =100$).
In this situation, it is expected that the presence of particle wakes modify the interactions between particles \citep{yin2007}. 
\begin{figure}[h!]
    \centering
    \includegraphics[height=0.21\textwidth]{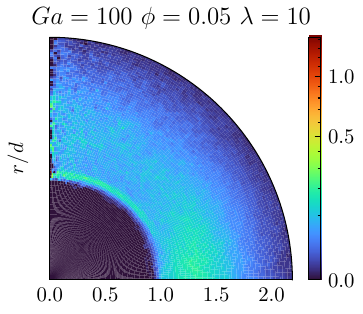}
    \includegraphics[height=0.21\textwidth]{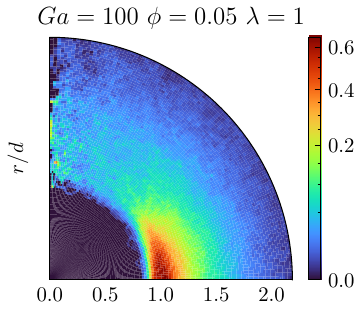}
    \includegraphics[height=0.21\textwidth]{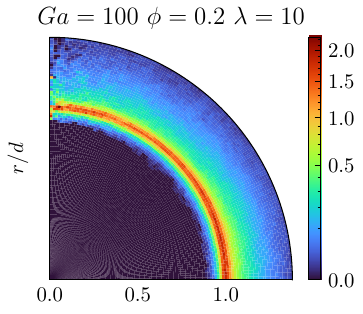}
    \includegraphics[height=0.21\textwidth]{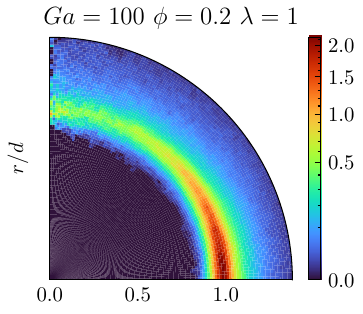}
    \caption{Histogram of the normalized function $P_\text{nst}$ at high inertia $Ga = 100$.
    The color map represents the values of the nearest pair distribution function. 
    The origin corresponds to the position of the \textit{\textit{test particle}}.
    The dimensionless radial and azimuthal coordinates, $|\textbf{r}|/d$ and $\theta$, correspond to the nearest neighbor position.
    The vertical direction corresponds to the flow direction, which is also the axis of symmetry for $P_\text{nst}$.
    (left) Low volume fraction cases $\phi=0.05$ for $\lambda = 1,10$.
    (right) High volume fraction cases $\phi=0.2$ for $\lambda = 1,10$.}
    \label{fig:Pnst_high_Ga}
\end{figure}
Anisotropy in \ref{fig:Pnst_high_Ga} for $\lambda=1$ is particularly striking compared to \ref{fig:Pnst_low_Ga}. 
In the former, a higher concentration of particles is identified at $\theta \approx 0$, as seen in \ref{fig:Pnst_high_Ga}. 
A higher concentration of $P_\text{nst}$ around $\theta \approx 0$ indicates the presence of horizontal rafts of particles. 
In this case, the microstructure is non-homogeneous and anisotropic; this situation is illustrated in \ref{fig:scheme_clusters} (\textit{Case 3: ``layers''}). 
As the \textit{Galileo} number ($Ga$) increases and for low values of the viscosity ratio ($\lambda$), the probability of having neighbors on the horizontal plane of the \textit{test particle} increases. 
This leads to an increase in the anisotropy of the microstructure, which is more pronounced for low volume fractions. 
In contrast, the high-viscosity drops display an isotropic distribution of the nearest particles around the \textit{test particle}. 
This observation suggests the presence of isotropic clustering of particles.



\begin{figure}[h!]
   \centering
   \includegraphics[width=0.4\textwidth]{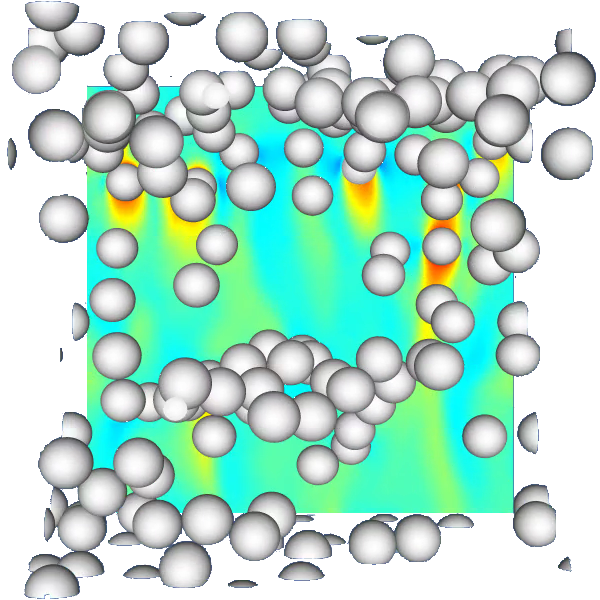}
   \includegraphics[width=0.4\textwidth]{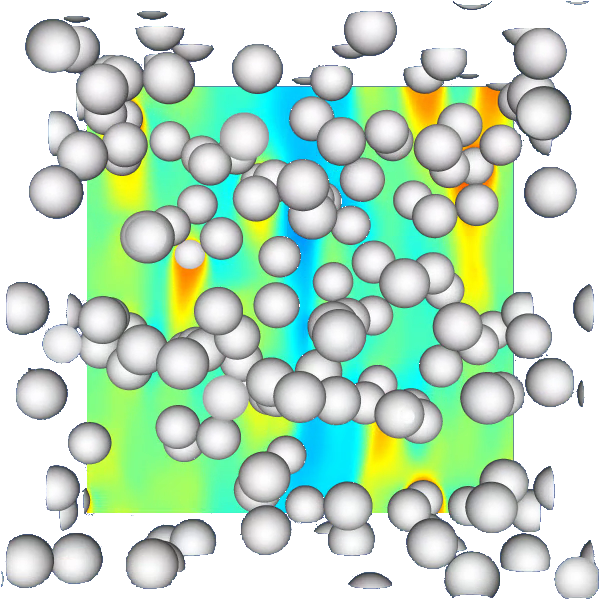}
   \caption{Snapshot of a simulation at $t^* = 150$ for $\phi=0.05$ and $Ga=100$.
   Color map : values of the vertical component of the velocity, field on the vertical plane defined by the equation $z=0$. 
   (left)  $\lambda = 1$.
   (right)  $\lambda = 10$.
   }
   \label{fig:images}
\end{figure}

To illustrate the impact of $\lambda$ on the microstructure, \ref{fig:images} displays snapshots of two DNS at $\phi = 0.05$ and $Ga = 100$. 
As predicted by $P_\text{nst}$, we observe layers and particles in close contact for $\lambda = 1$, contrasting with the seemingly more evenly dispersed microstructure for $\lambda = 10$.
In fact, for $\lambda = 10$, in \ref{fig:images} (right), we can still observe horizontal rafts of droplets or droplets rising side-by-side, but this effect is not as pronounced as for $\lambda = 1$. 
Drops with high viscosity ratios maintain a significant distance between other drops, which prevents the creation of structures such as droplet layers.
From the present analysis of $P_\text{nst}$ and the actual microstructure presented in \ref{fig:images} we can infer that the \textit{nearest particle statistics} is able to predict features in the microstructure such as layers and clusters.

\subsection{Nearest particle radial distribution function }

Although \ref{fig:Pnst_low_Ga} and \ref{fig:Pnst_high_Ga} give a good qualitative representation of the particle-pair azimuthal distribution, they fall short in delivering a quantitative depiction of the radial distribution.
Thus, in this section we investigate the value of the radial distribution function $P_\text{r-nst}$ defined in \ref{eq:P_r}. 
For a random isotropic distribution of hard spheres, it is possible to derive a theoretical prediction for $P_\text{nst}$ obtained in the vanishing volume fraction limit. 
Indeed, it is shown in \citet{zhang2021ensemble} that for a dilute random arrangement of particles $P_\text{nst}(r)$, reads as
\begin{equation}
    P_\text{nst}^{\phi \ll 1}(r) = n_p e^{- 8\phi\left[(r/d)^3-1\right]}.
    \label{eq:Pnst_dilute}
\end{equation}
It must be understood that this formula is accurate only at $\mathcal{O}(\phi)$; therefore, in most cases, it is not expected to be representative.
Additionally, \citet{torquato1990nearest} derived a radial distribution function for hard spheres at arbitrary volume fractions $\phi$. In our notation, this distribution can be written
\begin{equation}
    P_\text{nst}^\text{th}(r) = 
        n_p\left(e+\frac{f}{(r/d)} +\frac{g}{(r/d)^2}\right)
    e^{-\phi\left[8e\left((r/d)^3-1\right)+12 f\left((r/d)^2-1\right)+24g\left((r/d)-1\right)\right]}
    \label{eq:torquato}
\end{equation}
with, 
\begin{align*}
    && e= \frac{1+\phi}{(1-\phi)^3},
    && f= \frac{-\phi (3+\phi)}{2(1-\phi)^3},
    && g= \frac{\phi^2}{2(1-\phi)^3}.
\end{align*}
It should be noted that both \ref{eq:Pnst_dilute} and \ref{eq:torquato} are derived for hard spheres. 
In particular, for hard sphere $P_\text{nst}^\text{th} = 0$ for $r<d$ while in our case, particles might deform at contact, meaning that $P_\text{nst}$ is finite for certain $r<d$. 
However, using these theoretical probability density functions for comparative purposes remains valuable. 

\begin{figure}[h!]
    \centering
    \includegraphics[height=0.3\textwidth]{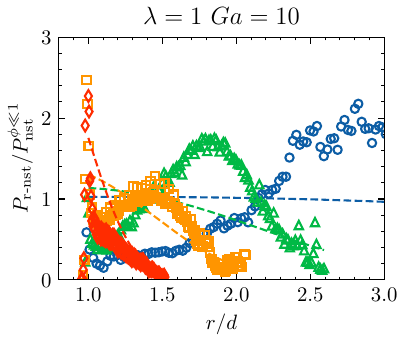}
    \includegraphics[height=0.3\textwidth]{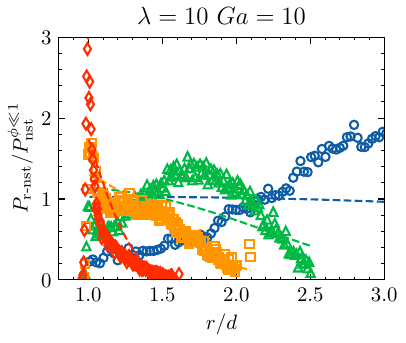}
    \includegraphics[height=0.3\textwidth]{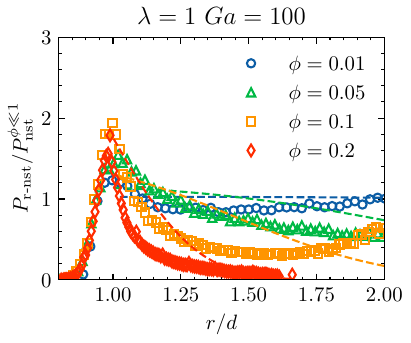}
    \includegraphics[height=0.3\textwidth]{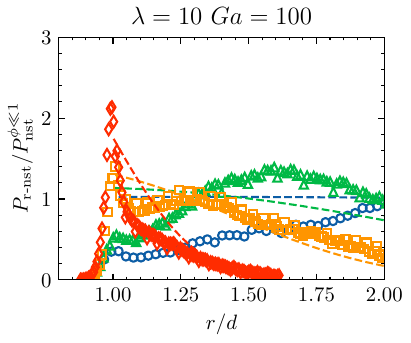}
    \caption{Radial probability density function divided by the theoretical distribution $P_\text{nst}^{\phi \ll 1}$ (\ref{eq:Pnst_dilute}), as functions of the dimensionless distance $r/d$, for $Ga = 100$, $Ga =10$, $\lambda = 1$ and $\lambda = 10$.
    (dashed lines) Theoretical prediction : $P_\text{nst}^{th}/P_\text{nst}^{\phi \ll 1}$. 
    (symbols) DNS results of $P_\text{r-nst}(r) / P_\text{nst}^{\phi \ll 1}$ with :
    ($\pmb\bigcirc$) $\phi = 0.01$; ($\pmb\triangle$) $ \phi = 0.05$; ($\pmb\square$) $\phi = 0.1$ ($\pmb\lozenge$) $\phi = 0.2$.
    For $r<d$ we arbitrarily set $P_\text{nst}^{\phi \ll 1} = 1$ so that $P_\text{r-nst}$ can be visualized. 
    }
    \label{fig:Pr}
\end{figure}
In \ref{fig:Pr}  we plot the radial distribution $P_\text{r-nst}$ as functions of the dimensionless distance $r/d$. 
We display two different viscosity ratios and multiple volume fractions. 
Let us first examine the low inertia cases displayed in \ref{fig:Pr} (top). 
It is seen that at $\phi = 0.01$ we have $P_\text{r-nst} < P_\text{nst}^{\phi\ll 1}$ for roughly $r/d < 2$.
This indicates that the likelihood of finding the nearest neighbor in the vicinity of the \textit{test particle} (i.e. $r/d<2$) is less than the random and dilute distribution.
In opposition, at $r/d>2$ the probability density $P_\text{r-nst}$ is higher than $P_\text{nst}^{\phi\ll 1}$ \eqref{eq:Pnst_dilute}.
In these cases ($\phi = 0.01$ and $\lambda = 10, 1$), the maximum of $P_\text{r-nst}$ is reached at $r/d \approx 3$, indicating that particles maintain on average a significant distance between each other. 
For comparison, the distance to the nearest neighbor in an ordered array of particles in a cubic lattice is  $\frac{1}{2}\left[4\pi/(\phi 3)\right]^{1/3} \approx 3.74$ at $\phi = 0.01$.
As $\phi$ increases, the maximum of $P_\text{r-nst}$ is found at lower values of $r/d$. 
At $\phi = 0.2$ the maximum of $P_\text{r-nst}$ is located at $r/d = 1$ which signifies that droplets are concentrated at the contact of the \textit{test particle}. 
At this stage, no differences between the cases $\lambda = 1$ and $\lambda = 10$ have been identified.

Let us now turn our attention to the inertial cases displayed on \ref{fig:Pr} (bottom). 
In the dilute regime ($\phi = 0.01$) and for $\lambda=1$, we observe that $P_\text{r-nst}$ is in complete agreement with the random and dilute distribution $P_\text{nst}^{\phi \ll 1}$ \eqref{eq:Pnst_dilute}.  
In contrast, when $\lambda = 10$ at the same $\phi$ and $Ga$, it becomes evident that fewer particles gather in close vicinity to the \textit{test particle}. 
For $\lambda = 10$ a behavior analogous to the low inertia scenario is observed. 
Indeed, the peak of $P_\text{r-nst}$ shifts leftward with increasing values of $\phi$, until the maximum of $P_\text{r-nst}$ is found at $r/d = 1$ for $\phi = 0.2$.
For $\lambda = 1$, we observe that $P_\text{r-nst}$ becomes progressively lower than the dilute and random distribution for increasing values of $\phi$ and moderate values of $r/d$. 
However, the maximum of $P_\text{r-nst}$ in these cases is located at $r/d = 1$ regardless of $\phi$. 
Thus, in the inertial and low viscous regime, the nearest neighbor is more likely to be located at near contact with the \textit{test particle}. 
Additionally, it is clear from \ref{fig:Pr} that the particles deform at contact, as witnessed by the non-vanishing value of $P_\text{r-nst}$ for $r/d<1$.
This phenomenon is particularly pronounced for the case $Ga = 100$ and $\lambda =1$. 
This is consistent with the high values of droplet aspect ratios reported in \ref{ap:deformation} \ref{fig:chi} for this case.
Interestingly, we note that in this case $P_\text{r-nst}$ exhibits the same values in the region $r/d <1$ for all values of $\phi$.

To conclude these observations, we note that both the radial and azimuthal distributions were affected by the inertial effects measured by the \textit{Galileo} number. 
The major effect of high inertia is the generation of strong anisotropy in the particle pair distribution and a more important concentration of neighboring particles at $r/d < 1$. 
Furthermore, the density at contact of the \textit{test particle} is higher for increasing volume fractions. 
The viscosity ratio strongly impacts the microstructure, but only at high $Ga$, whereas at low $Ga$, the change in viscosity ratio has no notable impact on $P_\text{nst}$. 
Lastly, note that \ref{eq:torquato} provides a significantly improved prediction of the radial distribution compared to $P_\text{nst}^{\phi \ll 1}$ \eqref{eq:Pnst_dilute} (see \ref{fig:Pr}). 
Indeed, the DNS results show that $P_\text{r-nst}$ follows trends similar to $P_\text{nst}^\text{th}$ as $\phi$ increases. 
Quantitative agreements are particularly observed for high inertia cases.

\subsection{Macroscopic modeling of the microstructure}
Until now, we have depicted the microstructure through 2D or 1D distributions. To provide a more succinct description, we propose a different method. 
Inspired by \citet{zhang2023evolution}, we will characterize the microstructure using the second moment of $P_\text{nst}$ with respect to $\mathbf{r}$, which is expressed as
\begin{equation}
    \textbf{R} =\frac{1}{n_p} 
    \int_{\mathbb{R}^3} 
    \textbf{rr} P_\text{nst}(\textbf{r}) d\textbf{r}.
    \label{eq:R}
\end{equation}
The tensor $\textbf{R}$ allows us to measure the mean-square distance between a particle and its nearest neighbor in the three dimensions of space.
It is worth noting that such a quantity is computable only because $\lim_{|\textbf{r}|\to \infty} P_\text{nst}(\textbf{r}) = 0$, which enables the integral of \ref{eq:R} to converge. 
This would not be the case for classical pair distributions, which is one of the main reasons for using the \textit{nearest particle statistics} framework. Since our objective is to measure the anisotropy of the microstructure, we are particularly interested in the deviatoric part of this tensor, namely
\begin{equation*}
    \textbf{A} = \textbf{R} - \frac{1}{3}  \text{tr}(\textbf{R}) \bm\delta.
\end{equation*}
Where $\bm\delta$ is the unit tensor and 
$\textbf{A}$ represents the likelihood of having a particle in a given direction compared to the mean radial square distance. 
Therefore, in an isotropic suspension, we have $A_{xx} = A_{yy} = 0$, where $y$ represents the coordinate aligned with gravity. 
In a situation like the one depicted in \ref{fig:images} (left), particles are more likely to have closer neighbors on the lateral sides rather than vertically positioned, hence  $A_{yy} < 0$ and $0 < A_{xx}$.

To facilitate the understanding of the subsequent findings, it is helpful to calculate the distance $r_m$. 
This distance represents the average distance to the closest neighboring particle within a random arrangement of dilute hard spheres. 
Through direct integration of \ref{eq:Pnst_dilute}, we get
\begin{align}
    \frac{r_m^2}{d^2}
    = 
    \frac{1}{d^2}
    \int_{\mathbb{R}^3} r^2 P_\text{nst}^{\phi \ll 1}(\textbf{r}) d\textbf{r} 
    =  \frac{\Gamma\left({{5}\over{3}} , 8\,\phi\right)}{8^{2/3}}\frac{e^{8\phi}}{\phi^{2/3}},
\end{align}
where $\Gamma(z,a) = \int_a^\infty t^{z-1} e^{-t} dt$ is the upper incomplete gamma function.
\begin{figure}
  \centering
  \includegraphics[height = 0.3\textwidth]{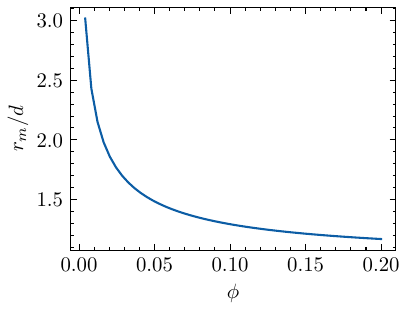}
  \caption{Dimensionless radial distance $r_m/d$ to the nearest neighbor for a random distribution.}
  \label{fig:agee}
\end{figure}
Since the upper incomplete gamma function is relatively uncommon, we have illustrated the dimensionless distance $r_m/d$ as a function of $\phi$ in \ref{fig:agee} to aid understanding. This figure clearly shows that $r_m$ decreases as $\phi$ increases.

\begin{figure}[h!]
    \centering
    \includegraphics[height=0.3\textwidth]{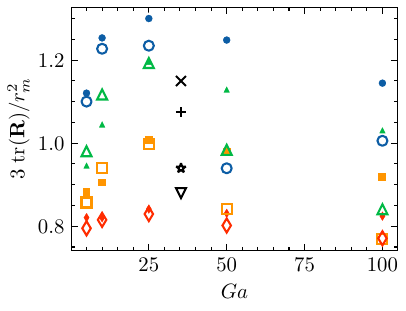}
    \includegraphics[height=0.3\textwidth]{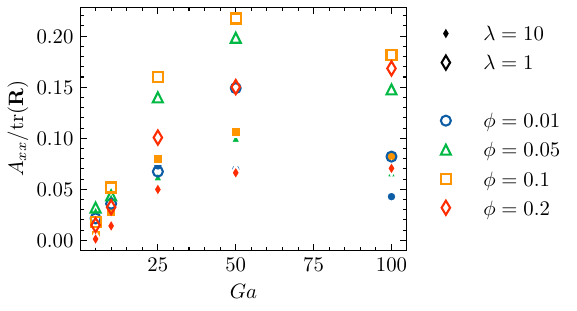}
    \caption{
        (left) Dimensionless trace of $\textbf{R}$ as a function of the volume fraction.
        (right) Horizontal components of the anisotropy tensor divided by the trace of $\textbf{R}$. 
    ($\pmb\bigcirc$) $\phi = 0.01$; ($\pmb\triangle$) $ \phi = 0.05$; ($\pmb\square$) $\phi = 0.1$ ($\pmb\lozenge$) $\phi = 0.2$.
    The hollow symbols correspond to $\lambda = 1$, the filled symbols to $\lambda = 10$.
    Black symbols represent the results of \citet{zhang2023evolution} for sedimenting spherical particles with $\phi = 0.016,0.056,0.134,0.262$ and $\zeta = 2.56$. 
    Corresponding to $\pmb\times,\pmb +, \pmb\star , \pmb\triangledown$, respectively.
    }
    \label{fig:A}
\end{figure}
In \ref{fig:A} (left), the dimensionless mean square distance between nearest neighbors, represented as $3\cdot\text{tr}(\textbf{R})/r_m^2 = \textbf{R}:\bm\delta/r_m^2$, is illustrated for all configurations explored in this study. 
The simulation marked by the symbol \textcolor{col1}{$\pmb\circ$} in \ref{fig:A} (left), corresponding to parameters $\lambda = 1$, $Ga = 100$, and $\phi = 0.01$, yields a value of $\bm\delta:\textbf{R}/r_m^2 \approx 1$. 
This observation agrees with the quasi-hard-sphere distribution previously reported in this case in \ref{fig:Pr} (left).
The dependence of the mean square distance on the volume fraction is also displayed on \ref{fig:A} (left). 
We observe a decrease in $\bm\delta:\textbf{R}/r_m^2$ as $\phi$ increases, indicating that particles tend to come closer to each other on average compared to a dilute random distribution of hard spheres. 
This observation, as highlighted by \citet{zhang2023evolution}, suggests the emergence of clusters when increasing the particle volume fraction. 
The dependence on the \textit{Galileo} number is non-monotonic. 
Indeed $\bm\delta:\textbf{R}/r_m^2$ increases until $Ga = 25$ and then decreases until $Ga = 100$. We also note that the distance to the nearest neighbor is not significantly affected by the viscosity ratio, particularly when dealing with high-volume fractions. 
In \ref{fig:A} (left), the symbols $\pmb\star$, $\pmb\times$, $\pmb +$, and $\pmb\triangledown$ depict the findings from the study conducted by \citet{zhang2023evolution} on the sedimentation of solid spheres in a liquid.
As observed, the value of $\textbf{R}:\bm\delta$ is, on average, closer to the mean $r_m^2$ than our simulations, but it maintains the same trend, i.e., clusters appear as the volume fraction increases.
 
\ref{fig:A} (right) illustrates the anisotropy of the microstructure. We can see on \ref{fig:A} (right) that we have $A_{xx} \ge 0$ for nearly all our cases, meaning that the emulsion is either isotropic ($A_{xx} = 0$), or exhibits a tendency towards a horizontal alignment of particles on average ($A_{xx} >0$). 
Additionally, $A_{xx}$ rises with $Ga$ up to $Ga = 50$, reaching a peak, and subsequently decreases until $Ga = 100$, although it remains significant.
In agreement with the remarks of the previous section, the value of $A_{xx}$ is greater for $\lambda = 1$ and lower for  $\lambda = 10$.
Although not obvious at first, we observe a non-monotonic trend with the volume fraction. $A_{xx}$ increases up to a peak value at $\phi = 0.1$ (indicated by \textcolor{col3}{$\pmb\square$} on \ref{fig:A} (right)), but then decreases for $\phi=0.2$ (shown by the \textcolor{col4}{$\pmb\lozenge$} symbols). 
This implies that at a certain volume fraction, around $\phi \approx 0.1$, higher $\phi$ makes the microstructure more isotropic, while at low volume fraction ($\phi < 0.1$), increasing $\phi$ favors the side-by-side configuration.
This phenomenon of isotropization at high $\phi$ has been reported in other studies such as in \citet{seyed2021sedimentation} for sedimentation of solid particles. 
However, at high \textit{Galileo} numbers, it seems that this effect is less pronounced.

To conclude, we classify the microstructure into four classes :
(1) The homogeneous microstructure.
(2) The non-homogeneous but isotropic microstructure with $\textbf{R}:\bm\delta > r_m^2$, or dispersed arrangement. 
(2 bis) The non-homogeneous but isotropic microstructure with $\textbf{R}:\bm\delta < r_m^2$, or clustering. 
(3) The non-homogeneous and non-isotropic microstructure or layering ($\textbf{A}\neq \textbf{0}$). 
Each of these types is characterized by specific values of $\textbf{R}$; they are reported in \ref{tab:microstructure}. 

\begin{figure}[h!]
    \centering
    \begin{tikzpicture}[scale=0.8]
        \node (img) at (0,0) {\includegraphics[height=5.5cm]{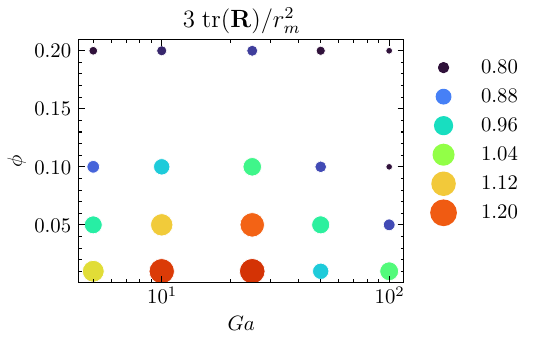}};
        \node (txt) at (-2,1) {Clustering};
        \node (txt) at (-1,-1.6) {Dispersed};
        \draw[dashed] ($(-1,-1.6) + (-10:3 and 2)$(P) arc
        (-10:155:3 and 2);
        \node (img) at (10.5,0) {\includegraphics[height=5.5cm]{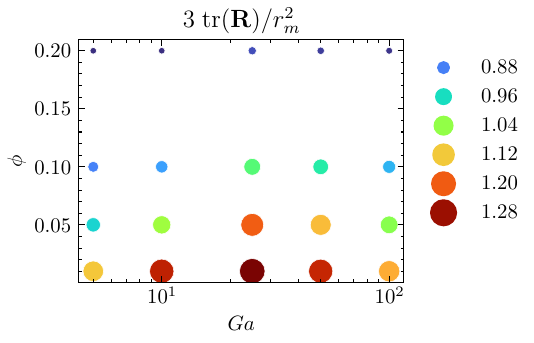}};
        \node (txt) at (8.5,1) {Clustering};
        \node (txt) at (10,-1.6) {Dispersed};
        \draw[dashed] ($(10,-2) + (-10:3 and 2)$(P) arc
        (-10:180:3 and 2);
    \end{tikzpicture}
    \begin{tikzpicture}[scale=0.8]
        \node (img) at (0,0) {\includegraphics[height=5.5cm]{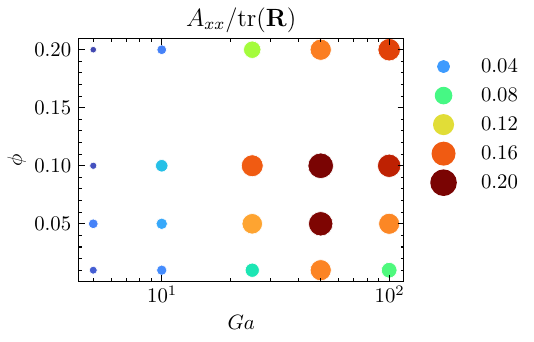}};
        \draw[dashed] (1.2,0.3) ellipse (1.5 and 2.5);
        \node (txt) at (1.2,1) {Anisotropic};
        \node (txt) at (-2,1) {Isotropic};

        \node (img) at (10.5,0) {\includegraphics[height=5.5cm]{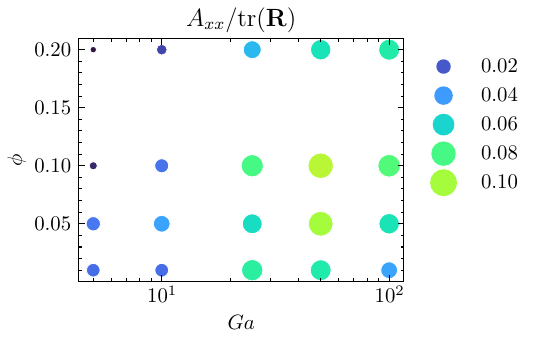}};
        \node (txt) at (8,1) {Isotropic};
    \end{tikzpicture}
    \caption{
        (top) Phase diagram of the dimensionless mean square distance to the nearest neighbor, $\bm\delta:\textbf{R}/r_m^2$.
        (bottom) Phase diagram of the dimensionless horizontal components of the anisotropy tensor, $A_{xx}/\text{tr}(\textbf{R})$.  
        (left) Iso-viscous emulsion $\lambda = 1$.
        (right) Viscous droplets $\lambda = 10$ }
    \label{fig:phase}
\end{figure}
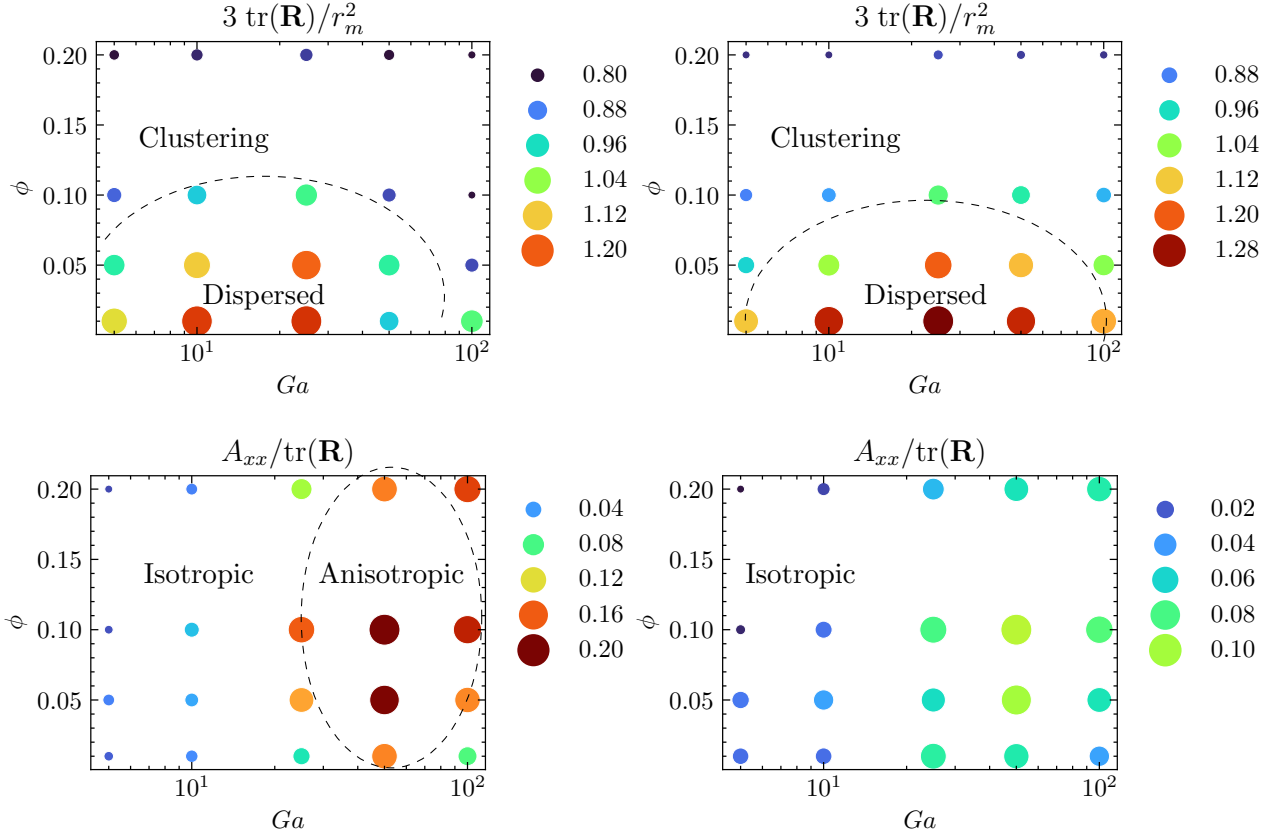

\begin{table}[h!]
    \caption{Microstructure classification}
    \label{tab:microstructure}
    \centering
    \begin{tabular}{|lccccc|} \hline
        Microstructure types & Homogeneous & Isotropic & \ref{fig:scheme_clusters} & $\textbf{R}:\bm\delta/r_m^2$ & $A_{xx}/tr(\textbf{R})$ \\
        Homogeneous & Yes & Yes &(\textit{Case 1}) & $ \approx 1$ & $\approx 0$ \\
        Dispersed &  No & Yes  &(\textit{Case 2}) & $ > 1$ & $\approx 0$ \\
        Clustering &  No & Yes  &(\textit{Case 2}) & $ < 1$ & $\approx 0$ \\
        Layering &    No & No  &(\textit{Case 4}) & $ - $ & $< 1$\\ \hline
    \end{tabular}
\end{table}
Additionally, to highlight the dependence of $\textbf{R}$ on $\phi$ and $Ga$, we display the values of $A_{xx}/tr(\textbf{R})$ and $\textbf{R}:\bm\delta/r_m^2$ in phase diagrams on \ref{fig:phase}.
We observe that the mean square particle distance compared to a random case decreases when increasing the volume fraction and is globally higher for viscous particles ($\lambda = 10$).
Meanwhile, the likelihood of finding the nearest neighboring particle on the horizontal is greater for $\lambda=1$ than $\lambda = 10$, and it is globally increasing with  $Ga$ and non-monotonic with $\phi$. 
We reach the configuration with the maximum anisotropy for both viscosity ratios at $Ga \approx 50$ and $\phi \approx 0.1$. 

\subsubsection*{Discussion}
While earlier studies primarily focused on bubbles or solid particles, it is reasonable to draw parallels between simulations with $\lambda = 1$ and $\lambda = 10$, corresponding to the former and latter scenarios, respectively. In \citet{bunner2002dynamics}, buoyant bubbles were simulated in a tri-periodic setup at Reynolds numbers around 10-30 for various volume fractions ($\phi$). The authors noted a tendency for the bubbles to align in horizontal pairs. 
This observation is consistent with what we observe in \ref{fig:phase} ($\lambda = 1$) since the anisotropy tensor is relatively high ($A_{xx} \approx 0.15$) for $Ga = 25$ (which corresponds roughly to $Re = 25$). 
Moreover, \citet{zhang2021direct} conducted Direct Numerical Simulations (DNS) of buoyant bubbly flows within tri-periodic domains. The Reynolds numbers varied from $Re=18$ to $22.8$ across different volume fractions, $\phi$, ranging from $0.05$ to $0.2$, with a fixed Galileo number, $Ga$, of $29$. They observed the presence of anisotropic clusters for $\phi > 0.1$, while noting their absence at lower volume fractions. This observation aligns with findings depicted in \ref{fig:phase} (left), where a small decrease in the value of $A_{xx}$ may be observed for the lowest $\phi$.

For solid particles at $Ga = 144$, it is observed in \citet{shajahan2023inertial} that vertical rafts of particles are formed in the dilute regime ($\phi \approx 0.02$). This phenomenon was attributed to a well-defined wake around individual particles, which effectively traps neighboring particles within the wake without causing repulsion toward the sides. 
In our case, we notice a greater concentration of particles in the vertical directions for $\lambda = 10$ compared to $\lambda = 1$ at $Ga = 100$, as evidenced by the smaller value of $A_{xx}$ in the former case. Although not immediately apparent, this phenomenon could be attributed to similar factors, namely, the wake of the viscous drop potentially inducing fewer instances of particles aligning side-by-side and more instances of vertically stable configurations. DNS at higher $Ga$ and $\lambda$ would be necessary to confirm or not the presence of the wake trapping effect. 
In the moderately dense regime,  $0.02 < \phi \le 0.1$  \citet{shajahan2023inertial} identified more configurations of particles situated side-by-side. 
As mentioned above, even if it is less pronounced than for $\lambda = 1$, we indeed observe that $A_{xx}$ is higher in these cases; see \ref{fig:phase} (right). 
Additionally, \citet{almeras2021statistics} carried out experiments of liquid-solid fluidized bed with spherical particles. 
Their Reynolds numbers range between $150\leq Re \leq 360$ depending on the volume fraction $0.14 \leq \phi \leq 0.42$.
It is observed that particles are most concentrated on the horizontal plane of the reference particle when $\phi = 0.14$.
If the $\lambda = 10$ cases follow the same trend, it is reasonable to expect that the probability of horizontal configurations, already predominant at $Ga =100$, will continue to increase for higher \textit{Galileo} at $\phi  \approx 0.1$.

We would like to end this comparison with the literature with the study of \citet{yin2008lattice} which compares the microstructure of suspensions of rising bubbles with suspensions of sedimenting solid particles.
The study encompasses two Reynolds numbers ($Re = 5, 20$) and two volume fractions ($\phi = 0.05, 0.2$). Their findings reveal that: 
\enquote{    
     microstructure in bubble
    suspensions is more anisotropic and inhomogeneous than
    solid particle suspensions of the same volume fraction and
    \textit{Reynolds} number.    
}. 
While our analysis focuses on emulsions with varying droplet viscosities rather than bubbles and suspended solid particles, our findings regarding the flow anisotropy align with the study of \citet{yin2008lattice}.


%% file: microstructure/Conclusion.tex
In this study, we provided a numerical framework to perform statistically steady simulations of rising emulsions with non-coalescing droplets. 
We then provided quantitative results to measure the microstructure geometry.
Three key points can summarize the major advances:
\begin{enumerate}
    \item We developed an optimized multi-VoF method within the \texttt{Basilisk} flow solver. 
    It enabled us to avoid coalescence between an arbitrary number of droplets while keeping the number of tracers inferior or equal to $7$. 
    Additionally, we showed that the multi-VoF method is capable of capturing the physics of interfacial interactions despite the coarse mesh definition at the scale of the film between the droplets, see \ref{ap:validation} (\textit{Case 2}). 
    This enabled us to perform massive DNS calculations of rising mono-disperse emulsion with various $\lambda$, $\phi$, and $Ga$ for long durations.
    \item We made use of the recent \textit{nearest particle statistic} framework of \citet{zhang2023evolution} to introduce an objective and concise way to measure the microstructure via the nearest particle pair density function $P_\text{nst}(\textbf{x},\textbf{r},t)$. 
    By looking at the pair distribution function, we could show qualitatively the influence of $Ga$, $\lambda$, and $\phi$ on the microstructure.
    Especially it was found that 
    (1) At low $Ga$ isotropic clusters (see \ref{fig:scheme_clusters}(\textit{Case 2})) appear with increasing $\phi$. 
    (2) Non-isotropic clusters (see \ref{fig:scheme_clusters}(\textit{Case 3})) are more likely to form for high \textit{Galileo} number.
    (3) The viscosity ratio $\lambda$ has an important impact on the microstructure: more clusters and layers are formed for $\lambda = 1$ than for $\lambda = 10$. 
    Even fewer clusters are observed for solid particle suspensions. 
    \item Following \citet{zhang2023evolution}, we show that the microstructure is well described by the second moment of the probability density, noted $\textbf{R}$. 
    This constitutes the central finding of this work. Indeed, we provided evidence that $\textbf{R}$ is a reliable and objective way to measure the microstructure.
    As predicted by \citet{zhang2023evolution} its trace measures the mean square distance between the nearest particles, ultimately a small trace witnesses of the presence of packed particles pair or clusters.
    Its anisotropic part indicates the presence of layers or side-by-side particle pairs in the flow. 
    With a reasonable estimation of the values of $\text{tr}(\textbf{R})$ and $A_{xx}$, one can predict the presence of side-by-side arrangements or clusters.
\end{enumerate}

We believe that the \textit{nearest particle statistics} framework is powerful enough to model the microstructure of multiphase-flow as it provides an objective and concise way to measure its pair distributions. 
Additionally, its ability to be included in a kinetic theory \citep{zhang2023evolution}, in opposition to other methods,  such as the Voronoi cell volume approach \citep{senthil2005voronoi}, makes it promising. As a perspective, based on the value of $\textbf{R}$, an approximation of the distribution $P_\text{nst}$ could be reconstructed by assuming a particular functional form with two degrees of freedom.
Each degree of freedom would correspond to the scalars $\text{tr}(\textbf{R})$ and $A_{xx}$.
This approximated distribution could then be used to take in account the various microstructures in some theoretical problems involving the theoretical pair distribution function (\citet{batchelor1972sedimentation}, \citet{hinch1977averaged,wang1999longitudinal}, and \citet{zhang2021ensemble}).

This work focused solely on the geometry of the microstructure in the steady state regimes. 
However, an important yet unexplored aspect of this problem is related to the timescales that drive the formation or evolution of the microstructure. 
In other words, what time does it take for the microstructure to reach its steady state? 
To determine how $\textbf{R}$ evolves in time and space, one can make use of the transport equation of $P_\text{nst}$ which directly yields a conservation equation for $\textbf{R}$. 
Then, this equation will eventually help us understand the kinematic of $\textbf{R}$ and, therefore, the kinematic of the microstructure. 
This will be investigated in a future study.

\section*{Acknowledgement}

The computational power of  \textit{TGCC - tr\`es grand centre de calcul du CEA} is greatly appreciated. 
\section*{Data availability}

All the data presented in this study are available upon request to the author. 
The buoyant emulsion simulations can be reproduced using the basilisk \texttt{.c} file \url{http://basilisk.fr/sandbox/fintzin/Rising-suspension/RS.c}, and following the instructions herein.

%% file: microstructure/AgeAP.tex
\section{Deformation of the drops}
\label{ap:deformation} 

Bubbles or droplets with a high Bond number are known to undergo significant deformations, and these deformations can significantly influence their rising velocity \citep{bunner2003effect,tripathi2014}. 
This appendix is dedicated to assessing the deformation of fluid inclusions within the range of dimensionless parameters prescribed in the main-body of the article. 
It is demonstrated that, across this entire range, the deformation remains small. 
Axisymmetric fluid inclusions are characterized by their aspect ratio, defined as the ratio of their cross-stream axis to the axis parallel to their velocity. 
However, for three-dimensional configurations involving deforming fluid inclusions, the definition of the aspect ratio becomes impractical due to the loss of symmetry in deformations \citep{bunner2003effect}. 
Consequently, to examine the shape of a given particle, we introduce its aspect ratio  
\begin{equation}
    \chi =  \sqrt{I_1 /I_2},
\end{equation}
with $I_1$ and $I_2$, the maximum and minimum eigenvalues of the particle inertia tensor defined as
\begin{equation*}
    \textbf{I}
    = \int_{V} \left[
        (\textbf{r}\cdot \textbf{r}) \bm\delta  - \textbf{rr}
        \right]
    d\textbf{x}.
\end{equation*}
Here, $\bm\delta$ is the unit tensor, $V$ the volume of a given particle and $\textbf{r} = \textbf{x} - \textbf{x}_i$, with $\textbf{x}_i$ is the center of mass of the fluid inclusion. 
Then, $\chi_p$ which is the averaged value of $\chi$ over all particles and timesteps, gives us a reliable way to compute the mean deformation of droplets. 
This definition has been used to characterize the deformation and orientation of rising bubbles \citep{bunner2003effect}. 
\begin{figure}[h!]
    \centering
    \includegraphics[height = 0.3\textwidth]{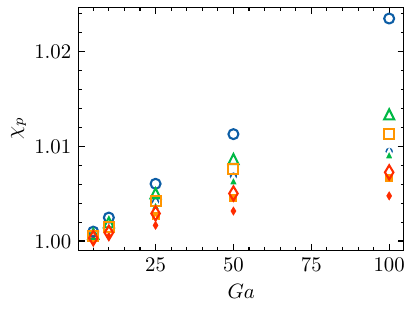}
    \caption{Mean aspect ratio of the droplets $\chi_p$, as a function of the \textit{Galileo} number, and the volume faction $\phi$,  for two different viscosity ratios.  
    The symbols correspond to different volume fraction ($\pmb\bigcirc$) $\phi = 0.01$; ($\pmb\triangle$) $ \phi = 0.05$; ($\pmb\square$) $\phi = 0.1$ ($\pmb\lozenge$) $\phi = 0.2$.
    The hollow symbols correspond to $\lambda = 1$, the filled symbols to $\lambda = 10$.
    }
    \label{fig:chi}
\end{figure}
\ref{fig:chi} displays the values of $\chi_p$ as a function of $Ga$.
As depicted in this figure, it is evident that droplets undergo flattening as $Ga$ increases and the drop volume fraction decreases. 
Additionally, it appears that, for lower viscosity ratios, the droplets exhibit a lower aspect ratio compared to more viscous drops.
In the low $Ga$ range, the average aspect ratio $\chi_p$ tends towards $1$, implying that in this regime droplets are mainly spherical. 
In summary, we observe a maximum $\chi_p$ value of $1.02$, which implies a deviation from a spherical shape of about $2\%$.

%% file: microstructure/validation.tex
The \texttt{Basilisk} code has been validated numerous times in previous numerical studies. 
Especially, we can cite the recent studies of \citet{innocenti2020direct} and \citet{hidman2023assessing} which both performed DNS of rising suspensions of bubbles. 
Nevertheless, in this work we investigate specific statistical distributions,
and we make use of a multi-VoF method to avoid droplets coalescence, therefore a meticulous validation of the DNS is in order. 
We start by presenting a brief comparison with the reference DNS of \citet{esmaeeli1999direct}. 
Afterward we present a study focusing on the interface kinematics where we compare our DNS with the experimental results of \citet{mohamed2003drop} to show that the multi-VoF method indeed captures the physics of two colliding interfaces without resolving the flow within the separating film. 
Once the mesh and the physics are validated, a study on the convergence of the statistics is presented. 

\subsection*{Ordered array of buoyant bubbles}

From our knowledge, no simulations nor experimental results have been carried out for rising buoyant viscous droplets. 
Therefore, we reproduced instead the ordered array simulation of \citet{esmaeeli1999direct} with \texttt{Basilisk} to validate the mesh resolution of our DNS.  
It consists in a 3-D buoyant ordered rising array of bubbles. 
In our notation the flow parameters of the simulation read 
\begin{align*}
    \lambda = 10,
    && \zeta = 10,
    && Bo = 1.8,
    && Ga = 28.37,
    && \phi = 0.125.
\end{align*}
\begin{figure}[h!]
    \centering
    \includegraphics[height = 0.3\textwidth]{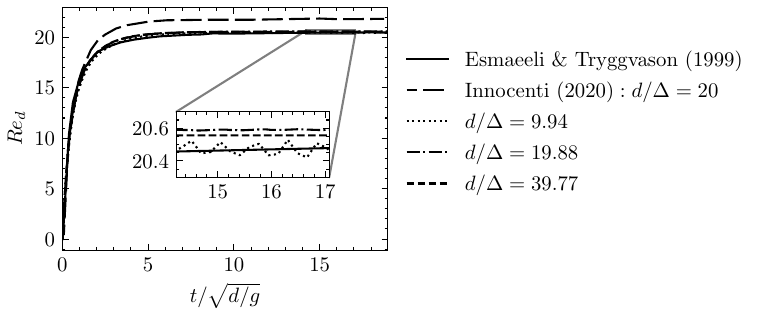}
    \caption{Time evolution of the Reynolds number based on the instantaneous volume averaged drift velocity, $Re_d(t) = \rho_fU _dd /\mu_f$, with $U_d(t)$ the drift velocity defined as $U_d = |\textbf{u}_d - \textbf{u}|$ with $\phi = 0.1256$, $\zeta =\mu_r =10$ and $Ga = 29.9$. $\textbf{u}_d$ and $\textbf{u}$ represent the volume-averaged velocities of the dispersed phase and the bulk, respectively, at time $t$.}
    \label{fig:ordered_array}
\end{figure}
\ref{fig:ordered_array} displays our numerical simulation against the original result of \citet{esmaeeli1999direct}.
We observe very good agreements between both studies for all mesh resolutions.
Additionally, we displayed the results of \citet{innocenti2020direct} for $d/\Delta = 20$ to point out a divergence with our results at the same mesh resolution.  
Both our simulations and the one of \citet{innocenti2020direct} have been carried out with the  \texttt{Basilisk} code. 
The cause of this difference is in fact due to a different method of interpolation used for the viscosity coefficient $\mu$. 
We used an arithmetic mean whereas \citet{innocenti2020direct} used an 
harmonic mean.
As a matter of fact in this regime the arithmetic mean, which will be used in this work, permits us to reach a faster convergence. 
Overall these results indicate that the criterion $d/\Delta = 20$ seems sufficient.

\subsection*{Drop impact on a liquid-liquid interface}

In this section we investigate in more detail the physics behind the multi-VoF method. 
We need to verify if we accurately capture the physics of the droplets interfaces despite the fact that we do not resolve accurately the film between two droplets. 
Following \citet{balcazar2015multiple} we reproduced the experiment of drop impact on a liquid–liquid interface carried by \citet{mohamed2003drop} but with the \texttt{Basilisk} code. 
This experiment consists in letting a drop fall into a pool of the same fluid as the drop. 
All along the experiment the interfaces of the droplets and the pool are tracked. 
In our notation the dimensionless parameters read 
\begin{align*}
    Ga = 71.02 
    && Bo = 6.40
    && \lambda = 0.33
    && \zeta = 1.189
\end{align*}
Following \citet{mohamed2003drop} we defined the dimensionless time $t / t_i = t U_i(t) /d$ where $U_i(t)$ is droplet velocity at $t<0$ and where $t=0$ is the time of impact. 
Regarding the geometry of the problem we sketched in \ref{fig:schemeLong} the initial position of the droplet in the computational domain.
Additionally, we display on \ref{fig:schemeLong} a snapshot of the numerical domain were we see the drop colliding the pool interface.
The drop and the pool do not merge since we use the multi-VoF method. 
Note that in the experiment the drop does not merge with the pool either.
This enables us to represent with the DNS a physical situation where the interfaces do not coalesce, but where we use a grid resolution of $d/\Delta = 20$ which is of course not sufficient to resolve the flow inside the film. 
\begin{figure}[h!]
    \centering
    \begin{tikzpicture}[ultra thick]
        \draw (0,0) rectangle (5,5);
        \draw[fill=gray!50] (0,0) rectangle (5,1.5);
        \draw[fill=gray!50] (2.5,3.5) circle (0.5);
        \draw[<->](0,-0.2) --++ (5,0)node[midway,below]{$L  = 10 d$};
        \draw[<->](-0.2,0) --++ (0,5)node[midway,left]{$L$};
        \draw[<->](5.2,0) --++ (0,1.5)node[midway,right]{$2.5 d$};
        \draw[<->](5.2,1.5) --++ (0,2)node[midway,right]{$5.5 d$};
        \draw[<->](5.2,3.5) --++ (0,1.5)node[midway,right]{$ 2d$};
        \draw[dashed,thin](2.2,3.5) --++ (2.9,0);
        \draw[dashed,thin](2.2,3.5) --++ (2.9,0);
        \draw[->](1,3.2) --++ (0,-0.7)node[midway,right]{$\textbf{g}$};
        \draw[<->](2,4.2) --++ (1,0)node[midway,above]{$d$};
        \draw[thin,dashed](2,3.3) --++ (0,1);
        \draw[thin,dashed](3,3.3) --++ (0,1);
        \node (a) at (4,2){$\rho_f, \mu_f$};
        \node (a) at (4,1){$\rho_d, \mu_d$};
    \end{tikzpicture}
    \includegraphics[height = 0.4\textwidth]{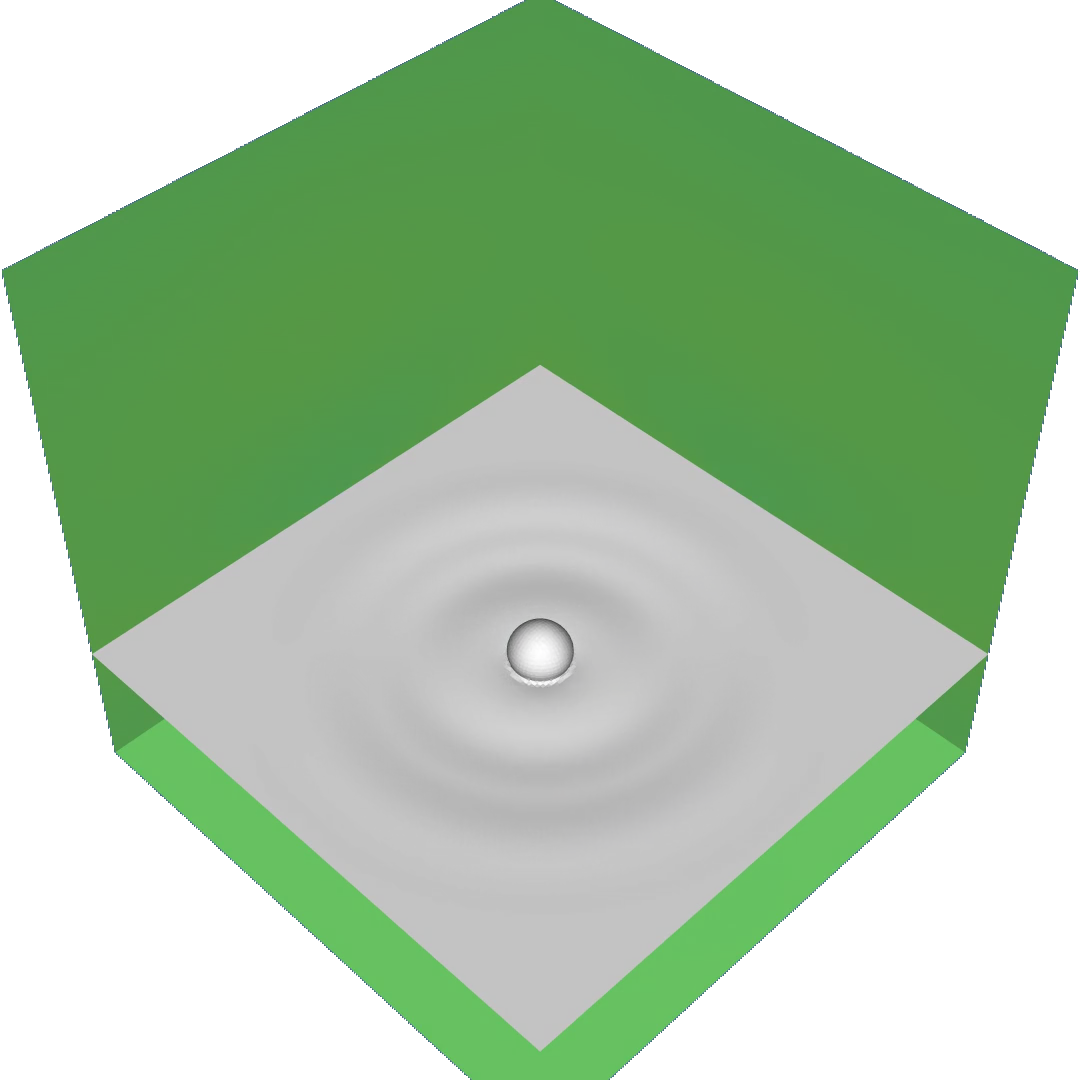}
    \caption{(left) Sketch of the computational set up at the initial time. 
    (right) Snapshot of the computational domain after the collision, with the pool interface represented in gray.
    The background color represents the velocity field magnitude, which is undisturbed, indicating a large enough domain. }
    \label{fig:schemeLong}
\end{figure}
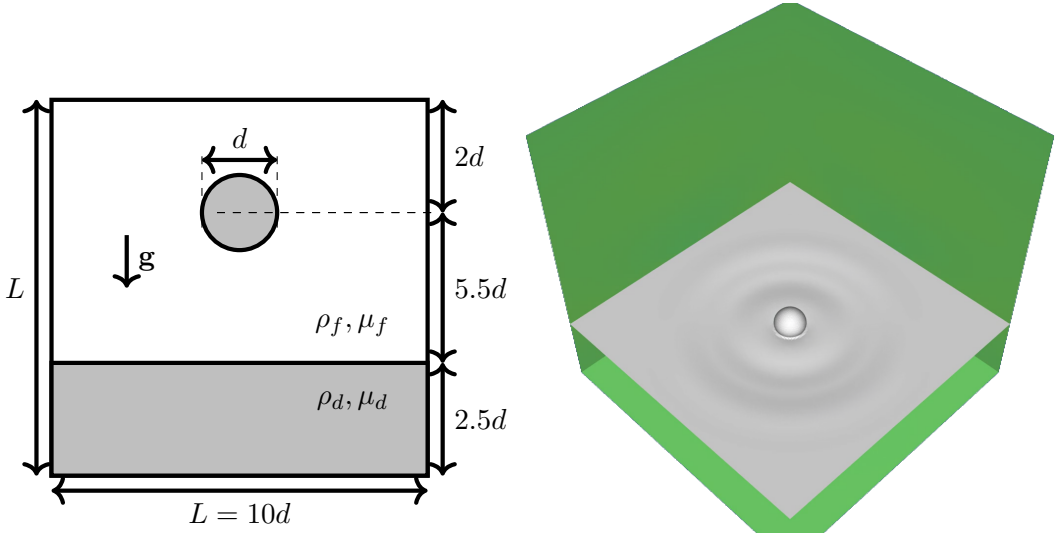
\begin{figure}[h!]
    \centering
    \includegraphics[height = 0.3\textwidth]{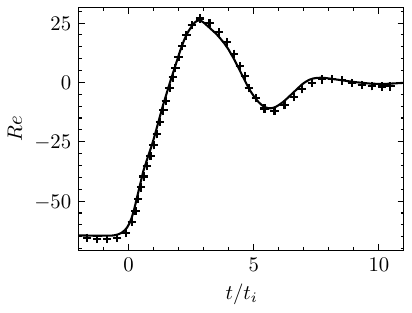}
    \includegraphics[height = 0.3\textwidth]{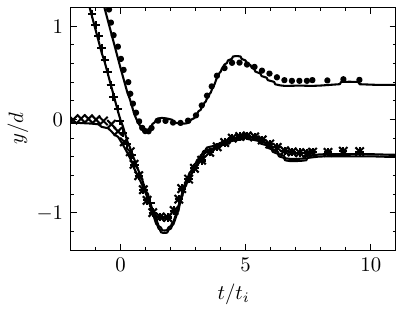}
    \caption{(left) Time evolution of the Reynolds number based on the droplet velocity, $Re(t) = \rho_fU d /\mu_f$ as a function of the dimensionless time, (+) numerical results of  \citet{balcazar2015multiple} (right)  position of the interfaces, ($\bullet$) top droplet surface, ($+$) bottom droplet surface, (x) pool surface. (Symbols) Experimental results of \citet{mohamed2003drop} (solid line) present numerical simulations with $d/\Delta = 20$. }
    \label{fig:resultslong}
\end{figure}
\ref{fig:resultslong} represents the comparison between our results and the experiment of \citet{mohamed2003drop} (right) and the numerical simulation of \citet{balcazar2015multiple} (left). 
The time-dependent Reynolds number as well as the interfaces positions are shown to closely match both the numerical and experiential results. 
From the very good agreement we conclude that the kinematics are well-represented even during contact for a mesh resolution of $d/\Delta = 20$.

\subsection*{Mesh independence and statistical convergence for random arrays of drops}

Even though the aforementioned studies carried validations of the \texttt{Basilisk} code for rising droplets or bubbles, almost all of them considered isolated droplets or bubbles as the only validation case. 
To the author's knowledge, to this date no published study has presented a mesh independence study for random arrays of droplets or bubbles of this scale. 
As particle interactions and higher \textit{Galileo} numbers may be more challenging to model, it is primordial to investigate the mesh independence of the DNS that are carried in this work. 
In this objective we performed a DNS of a random array of $N_b=125$ droplets, with the following parameters
\begin{align*}
    \lambda = 10,
    && \zeta = 1.11,
    && Bo = 0.2,
    && Ga = 100,
    && \phi = 0.1,
    && N_b =125,
\end{align*}
with mesh resolutions of $d/\Delta = 5,\; 10,\; 18,\; 37$. 
This set of parameters have been selected following these arguments :
A viscosity ratio $\lambda = 10$ induces more vorticity at the droplets interfaces in contrast with the $\lambda = 1$ cases. 
For high inertia regimes ($Ga = 100$) the boundary layers at the droplet interfaces require the fine grid to be resolved compared to the low inertia cases. 
At $\phi = 0.1$, numerous interactions of droplets are present, implying that the good modeling of the liquid films between interfaces becomes predominant on the overall hydrodynamic, this also requires a good mesh resolution. 
For these reasons, we suppose that this case might require the finest grid among all other cases presented in this study. 
Based on this remark we can assume that if this case is mesh independent, then all cases from \ref{tab:simulations} are equally validated. 

Let us first verify the independence of the drift velocity on the mesh resolution. 
\begin{figure}[h!]
    \centering
    \includegraphics[height = 0.3\textwidth]{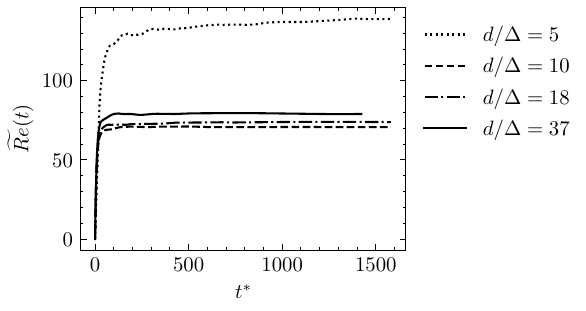}
    \caption{
        Running average of the Reynolds number based on the instantaneous volume-averaged relative velocity, $Re(t) = \rho_fUd /\mu_f$, with $U(t) = |\textbf{u}_d - \textbf{u}_c|$ for $\phi = 0.1$, $Ga=100$ and $\lambda =10$. $\textbf{u}_d$ and $\textbf{u}_c$ represent the volume-averaged velocities of the dispersed phase and the continuous phase, respectively, at time $t$.
        In the legend we display the value of the mesh resolution. 
    }
    \label{fig:Re}
\end{figure}
In \ref{fig:Re} we display the running-averaged drift velocity as a function of time, for four mesh resolutions. 
The results are not as independent of the mesh resolution as the ordered array validation presented above. 
Indeed, we observe a difference of the rising Reynolds number of about $5\%$ between the $d/\Delta = 18$ and $d/\Delta = 37$ cases which is notable.
We recall that this $5\%$ error will eventually be lower for all other cases. 
The good agreement between the case  $d/\Delta = 10$ and $d/\Delta = 18$ is partially fortuitous.